\documentstyle[emulateapj,epsf]{article}

\def\url#1{{\ttfamily\def\/{/\discretionary{}{}{}}#1}}
\def\simless{\mathbin{\lower 3pt\hbox
   {$\rlap{\raise 5pt\hbox{$\char'074$}}\mathchar"7218$}}} 
\def\simgreat{\mathbin{\lower 3pt\hbox
   {$\rlap{\raise 5pt\hbox{$\char'076$}}\mathchar"7218$}}} 

\def\kms{{\rm\,km\,s^{-1}}}

\def\gcm3{{\rm g\,\, cm^{-3}}}
\def\msun{M_\odot}
\def\pc3{{\rm pc}^{-3}}

\def\bh{{\rm BH}}

\def\B{{\rm B}}
\def\Mbh{M_{\rm BH}}

\def\mbh{m_{\rm BH}}
\def\mbhf{m_{\rm BH,f}}
\def\MBH{ {\cal{M}_{\rm BH}}}

\def\T{{\rm T}}

\def\be{\begin{equation}}  
\def\ee{\end{equation}}  
\def\baray{\begin{eqnarray}}
\def\earay{\end{eqnarray}}  
\def\psra{[PSR-A,CO]~}
\def\psa{\rm {PSR-A}}
\def\ns{[NS,CO]~}
\def\bhbh{[BH,BH]~}
\def\bhcusp{[BH,*]~}
\def\bhns{[BH,NS]~}

\begin{document}

\title {Probing the presence of a single or binary black hole 
in the globular cluster NGC 6752 with pulsar dynamics}

\author{Monica Colpi\altaffilmark{1}, Michela Mapelli\altaffilmark{1}, \& 
Andrea Possenti\altaffilmark{2,3}}

\affil{\altaffilmark{1}Dipartimento di Fisica G. Occhialini,\break
  Universit\`a di Milano Bicocca, Piazza della Scienza 3. I-20126 Milano, Italy}  

\affil{\altaffilmark{2}INAF-Osservatorio Astronomico di Cagliari,\break
Loc. Poggio dei Pini, Strada 54, I--09012 Capoterra, Italy}

\affil{\altaffilmark{3}INAF-Osservatorio Astronomico di Bologna,\break
Via Ranzani 1, I--40127 Bologna, Italy}

\begin{abstract}
The five millisecond pulsars that inhabit NGC 6752 display locations
or accelerations that are quite unusual compared to all other pulsars
known in globular clusters.  In particular PSR-A, a binary pulsar,
lives in the cluster halo, while PSR-B and PSR-E, located in the core,
show remarkably high negative spin derivatives.  This is suggestive
that some uncommon dynamical process is at play in the cluster core
that we attribute to the presence of a massive perturber.  We here
investigate whether a single intermediate-mass black hole, lying on
the extrapolation of the mass $\cal {M}_{\rm BH}$ versus $\sigma$
relation observed in galaxy spheroids, or a less massive binary black
hole could play the requested role. To this purpose we simulated
binary-binary encounters involving PSR-A, its companion star, and the
black hole(s).  A ``stellar-mass binary'' black hole of
($50\msun,10\msun$) can imprint the right thrust to propel PSR-A in
the halo during a flyby.  The flyby is gentle and does not alter the
internal properties of the binary pulsar.  An ``intermediate-mass
binary'' black hole of ($ 200\msun,10\msun$) tends to impart a recoil
speed larger than the escape speed: It can release PSR-A on the right
orbit if its separation is wide.  A ``single intermediate-mass'' black
hole of mass $\MBH\simgreat 500\msun$ may have ejected PSR-A at the
periphery of NGC 6752 in a close dynamical encounter involving the
binary pulsar, the black hole and a star belonging to its cusp.  The
encounter gives correct speeds but alters significantly the
eccentricity of the impinging binary, so that it must occur before the
neutron star of PSR-A is recycled via accretion torques.  The
influence of an intermediate-mass binary black hole on the
acceleration of the two core pulsars is studied, and the ejection of
stars by the binary is shortly explored.  In inspecting our close
4-body encounters, we have found that a single or binary black hole
may attract on a long-term stable orbit a millisecond pulsar. Timing
measurements on the captured ``satellite'' pulsar, either member of a
hierarchical triple or of the cusp, could unambiguously unveil the
presence of a black hole(s) in the core of a globular cluster.
\end{abstract}


\section{Introduction}

Five millisecond pulsars have been recently discovered in 
NGC 6752 displaying 
unexpected characteristics (D'Amico et al. 2002).  PSR-A, a canonical
recycled binary pulsar, holds the record of being the farthest
millisecond pulsar ever observed from the gravitational center 
of a globular cluster, at a distance of $\approx$ 3.3 half mass
radii.  PSR-C, an isolated pulsar, ranks second in the list of the
most offset pulsars, being at a distance of 1.4 half mass radii from
the center\footnote {D'Amico et al. (2002) considered the possibility of
a  chance superposition of a galactic field millisecond pulsar  
inside  the observed projected distance from the center of NGC6752, inferring a
probability $<10^{-4}$  for each of the two offset sources.}.
PSR-B, PSR-E, and PSR-D are located instead within the
cluster core and are single; PSR-B and PSR-E have remarkably high {\it
negative} spin derivatives while 
PSR-D has a  {\it positive} $\dot P$, one of  
the highest measured among the 
globular cluster pulsars.  If the 
negative spin derivatives of PSR-B and PSR-E 
are  ascribed to the overall effect of
the cluster gravitational potential, this would result in a 
central projected  mass to light ratio $M/L_V\sim 6-7$  
(Ferraro et
al. 2003a).
This high ratio may imply the presence
of $\approx 1,000-2,000 \msun$ of under-luminous matter enclosed
within the central 0.08 pc of the cluster (Ferraro et al. 2003a),
perhaps in the form of a single intermediate-mass black hole and/or
in the form of collapsed stellar remnants. In
this scenario even the very high positive $\dot{P}$ of PSR-D 
could be explained as due to the line-of-sight gravitational
pull of this
unseen matter. However, as pulsars spin down
due to rotational energy losses, the period derivative
of PSR-D could
also be intrinsic. Alternatively, one  could argue that
the negative values of $\dot{P}$ of PSR-B and PSR-E would result from
the gravitational pull of some {\it local perturber}, such as a nearby
passing star or even a more massive exotic objects (Ferraro et
al. 2003a).  
The millisecond pulsars in
NGC 6752 are peculiar in their location or
acceleration, and the combination of these facts strongly suggests
the occurrence of an {\it uncommon dynamics} in the core and halo of NGC 6752.
It is our aim to address this issue here in detail.

In a previous paper (Colpi, Possenti, \& Gualandris 2002, CPG
hereafter) we explored a number of roots
for the origin of PSR-A: PSR-A  may have originated
from a primordial binary,  born  either in the halo or in the core;
but a careful analysis (based on considerations on 
characteristic lifetimes and neutron star natal kicks) 
led us to discard both these two
hypotheses (see CPG). A third, involving a 3-body scattering
or exchange event off core stars, was also
rejected given the tight constraints imposed
by the binary nature of PSR-A (CPG).  We thus were led to conjecture
that a more massive target, such as 
{\it a binary of two black holes} with masses in the range
$\approx 10-100\msun$ could have provided, in a 4-body scattering event,
the necessary thrust to propel PSR-A into its current halo orbit, at
an acceptable event rate (CPG).

PSR-A may just signal the presence
of a black hole binary in NGC 6752, but this is somewhat
puzzling.
Black hole binaries are
expected to form  in rich star clusters; but along the course of
evolution they are expected to exit 
their parent cluster (Kulkarni, Hut, \&
McMillan 1993;
Sigurdsson \& Hernquist 1993).
The black holes (relics of the most massive stars)  
tend to pair 
with  other black holes in binaries 
as soon as they segregate in the cluster core by
dynamical friction (Kulkarni, Hut, \& McMillan 1993).  Some binary may
rapidly merge emitting gravitational waves (Benacquista 1999; Miller 2002)
creating a more massive black hole (Miller \& Hamilton 2002), but
most/many
leave the cluster since close (3- or 4-body) dynamical encounters 
among the black holes eject
them, single or in binaries, due to recoil (Sigurdsson
\& Hernquist 1993; Kulkarni, Hut, \& McMillan 1993). 
Current N-body simulations
(Portegies Zwart \& McMillan 2000) 
expect no black hole or one black hole binary to remain in the cluster.
Thus, if the interpretation of PSR-A is correct, we may provide first dynamical 
evidence of the ``only'' black hole binary that avoided the escape.
A goal of this paper is to further explore this possibility
narrowing the range
of masses of the two black holes from pulsar dynamics.

There is an alternative
root for the origin of 
the unusual location of PSR-A that we wish to explore here in 
connection to
the  high $M/L_V$ ratio, and so to the potential presence of  
under-luminous matter in the cluster core:  
the 
ejection of PSR-A into the halo of NGC 6752 by a dynamical encounter
with {\it a central intermediate-mass black hole}.    
According to an old
suggestion by Frank \& Rees (1976) stars in the {\it cusp} of a
massive black hole can eject other stars plunging inside 
(Lin \& Tremaine 1980).  A flyby involving a
central black hole, a bound star orbiting around it, and the binary
pulsar, may imprint to PSR-A the right recoil speed to climb the potential
well of the cluster and reach the halo.

Single black holes of intermediate mass 
($\MBH \simgreat 
500\msun $) may inhabit the center of globular clusters; their formation root
can involve the runaway growth of a super-massive star
through collisions of heavy 
stars in the young cluster 
(Portegies Zwart \& McMillan 2002) 
or the occurrence of 
repeated mergers among  compact objects
(Miller \& Hamilton 2002; see van der Marel
2003 and Miller 2003 for a review).
Recently,  
HST/STIS observations of the globular cluster G1 in M31
(Gebhardt et al. 2002) and of M15 in the Milky Way (Gerssen et
al. 2002, 2003) have provided first clues for the presence of a single central
black hole.  Surprisingly, 
the two postulated black holes, of
mass $2.0^{+1.4}_{-0.8}\times 10^4 \msun$ in G1, and of 
$1.7^{+2.7}_{-1.7}
\times 10^3\msun$ in  M15, seem to lie just along
the extrapolation of the black hole mass versus 1D dispersion velocity
relation ($\cal
M_{\rm BH}-\sigma$)  obeyed by the super-massive black holes in
galaxy spheroids (Ferrarese \& Merritt 2000; Gebhardt et al. 2000;
Tremaine et al. 2002).
Here we assume that the 
hypothetical black hole in NGC 6752 just lies on the $\cal{M}_{\rm BH}-
\sigma$
relation.  
(Note that current evidence of a large black hole in G1 and M15 is  not yet
compelling:  Baumgardt et al. 2003a, and 2003b 
provide an alternative to the black hole hypothesis, i.e., 
a cluster of lower-mass collapsed objects.) 

In this paper, we simulate a series of binary-binary
encounters to study (i) the scattering of the binary
pulsar PSR-A off  a {\it stellar mass or intermediate-mass 
black hole binary}, and (ii)  
the plunge-in of  PSR-A 
inside {\it the cusp of a single intermediate-mass black
hole}.  
In $\S 2$ we select the mass spectrum of the two black holes
in the binary and choose the mass of the single intermediate-mass 
black hole.
In $\S 3$ we describe properties of the binaries, 
and identify the possible end-states
of binary-binary encounters
(flyby, ionization and exchange). 
The binary hosting the neutron star
is described in its post-recycling phase as well as in its pre-recycling state.
In $\S 4$
we extract from a series of binary-binary simulations 
post-encounter distributions of recoil velocities,
eccentricities and binary separations, 
running  more than $20,000$ encounters.
In $\S 5$ we examine the statistics of the encounters, while
in $\S 6$ we shortly explore exotic end-states. 
In $\S 7$, we model the random walk 
(moderated by dynamical friction) of the black hole binary induced by 
stellar collisions, and try to evaluate the local gravitational
pull that a black hole binary may exert 
on the core pulsars, inside NGC 6752. 
In $\S 8$ we explore  
the survival of a black hole binary in 
NGC 6752, 
mimicking mass growth and hardening by stellar encounters. 
In $\S 9,$ we outline the key findings and present our conclusions.

\section { Black hole masses}

The masses  of the two black holes in the binary, denoted 
$\Mbh$ (for the heavier)
and $\mbh$ (for the lighter),  are a free parameter (let
$M_\T=\Mbh+\mbh$ be the total mass of the binary). 
Constraints on their values derive mainly from 
limits on their 
ejection due to gravitational encounters and survival against coalescence 
by  emission of gravitational waves (GWs hereafter). 

In general, survival inside the cluster (followed by   
coalescence due to GWs) would win over ejection 
if the binary,  
at the separation $a_{\rm eq}$ where the time-scale for coalescence 
$\tau_{\rm GW}=
[5c^5a_{\rm eq}^4(1-e^2)^{7/2}]/[256 G^3\Mbh\mbh M_\T]$
becomes comparable to the collision time 
$\tau_{\bh}\sim \sigma_{\bh}/[n_{\bh}GM_\T\pi a_{\rm eq}],$
has a  binding energy $E_{\rm bin}=G\Mbh\mbh/2a_{\rm eq}$
smaller than  the minimum 
binding energy for expulsion, $E_{\rm ej}(V_{\rm es})$,
given an encounter with a field black hole
(here, $n_{\bh}$ and $\sigma_{\bh}$ are the density and 1D dispersion
velocity of the field black holes interacting with the binary, 
and $V_{\rm es}$ is the escape
speed from the core).
If $\mbhf$ is the mass of the field black hole 
impinging on the binary, momentum conservation implies 
a minimum energy for escape equal to 
\be
E_{\rm ej}(V_{\rm es})=\left ({1\over 2\xi_\bh}
\right )(M_\T+\mbhf)\left ({M_\T\over
\mbhf}\right )^{2}V^2_{\rm es},
\ee
where $\xi_\bh$ controls the relative energy exchange 
per scattering (see Table 1 and Table 3 for the values of  
$\xi_\bh$).
If $E_{\rm ej}>E_{\rm bin}$  at $a_{\rm eq},$ 
the binary avoids ejection during  its entire lifetime (when its separation
$a>a_{\rm eq}$), and  may eventually  merge producing
a more massive black hole when $a$ drops below $a_{\rm eq}.$
{ 
\vskip 0.2truecm \epsfxsize=9.truecm \epsfysize=9.truecm
\epsfbox{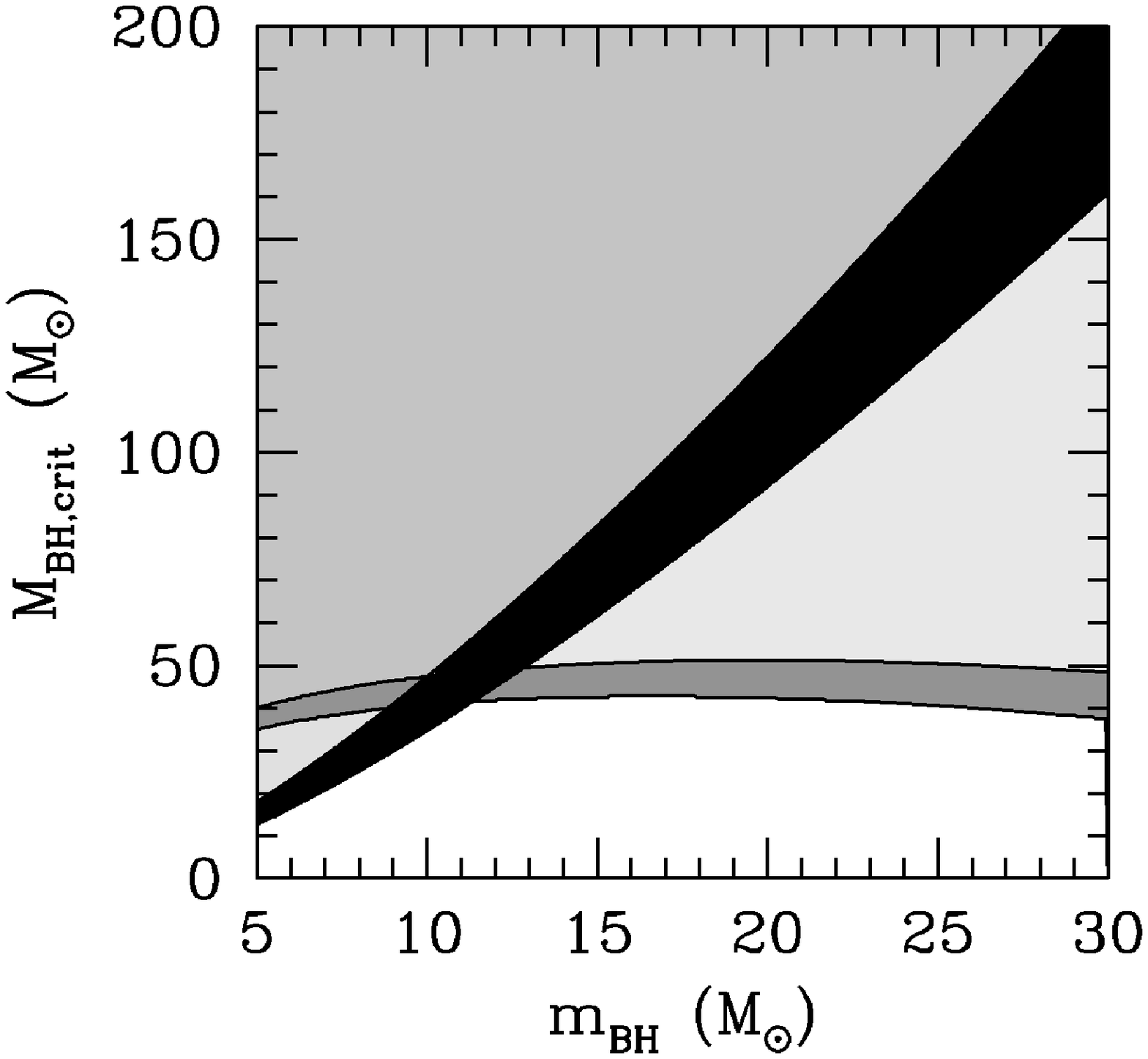} 
\figcaption[figure1.eps] 
{
\label{fig:fig1}
\footnotesize 
{
$M_{\rm BH,crit}$ against $\mbh$ (in units of $\msun$), for
a close dynamical encounter with a field black hole of mass $\mbhf=\mbh$
(black strip) and $\mbhf=10\msun$ (dark grey strip). The binary has
eccentricity $e=0.7.$ 
The escape 
speed is $V_{\rm es}=40\kms$ and 
$\xi_\bh=0.5$ (Miller \& Hamilton 2002). The background 
black hole density, and dispersion velocity are set equal to $10^6\pc3$ 
and $7\kms,$ respectively. In generating the strips
we introduced some scatter in $\xi_\bh,$  $V_{\rm es}$ and 
density, to bracket uncertainties.
The shadowed areas in light grey above each strip 
indicate the permitted values of
$\Mbh,$  consistent with binary black hole retention in the cluster. 
}
}
}
\vskip 0.2truecm


The 
condition $E_{\rm ej}(V_{\rm es})=E_{\rm bin}(a_{\rm eq})$
selects a critical mass $M_{\rm BH,crit}$ 
for the heavier black hole in the binary, for  fixed $\mbh,$
above which ejection by recoil off a field black hole is avoided.
In Figure~\ref{fig:fig1}, we plot $M_{\rm BH,crit}$ against $\mbh$, considering
a close dynamical encounter with a field black hole of mass $\mbhf.$
The black strip refers to a 
mass $\mbhf$ set equal to $\mbh,$ while the 
dark grey  strip refers to  scattering with a field black hole  
of fixed mass $\mbhf=10\msun.$  Above the two strips (shadowed light
grey areas  
depending on
$\mbhf$) binaries
with $\Mbh>M_{\rm BH,crit}$ remain in the cluster.

Figure~\ref{fig:fig1} shows clearly that black holes
with larger masses ($\simgreat 30 \msun$) 
and small mass ratios $\mbh/\Mbh$ are preferentially retained.
We have considered 
values of ($\Mbh/\msun,\mbh/\msun$)  equal to 
(10,10), (30,3),(50,10), and (200,10) for the binary black holes
in our simulations of binary-binary encounters; in one series of run 
we considered a (50,1.4) binary 
comprising a  black hole and a canonical neutron star
\footnote { $3\msun$ is the most general 
upper limit on the mass of a neutron star
that we considered as lower limit for the mass of the lighter black hole.}. 
The (10,10) binary is in the forbidden region;  
it is studied to set comparisons with heavier binaries, and because its 
mass is in the range of observed black hole masses 
(Bailyn et al. 1998).  Binaries with the heaviest black hole
below $100\msun$ will be considered as {\it stellar-mass black hole
binaries}. Those with  $\Mbh>100\msun$ will be referred to as
{\it intermediate-mass black hole binaries}. 

{ 
\vskip 0.2truecm \epsfxsize=8.truecm \epsfysize=8.truecm
\epsfbox{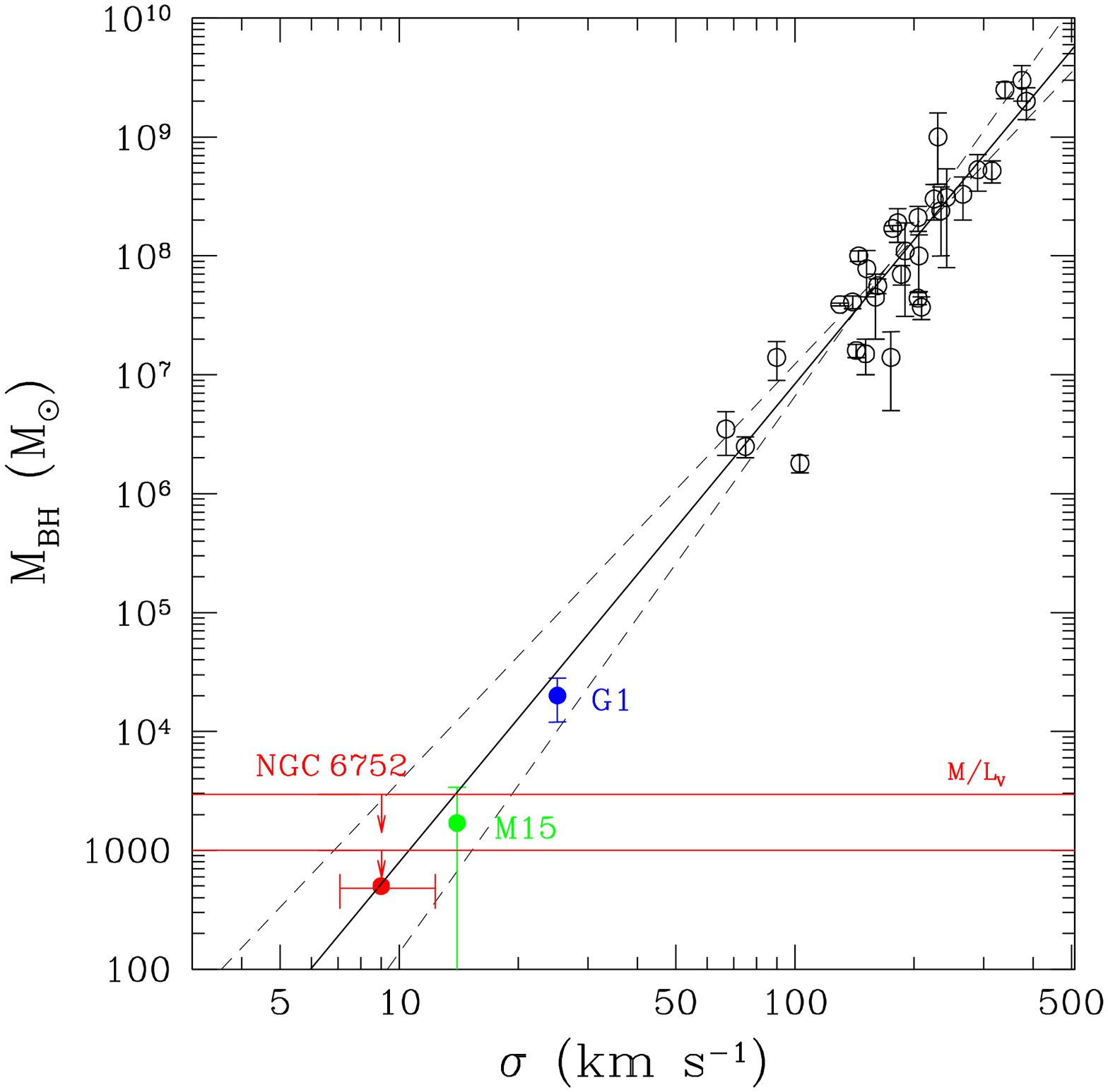}
\figcaption[figure2.eps] 
{
\label{fig:fig2}
\footnotesize 
{
$\MBH$ versus $\sigma$ relation for super-massive black holes
in galaxy bulges (Tremaine et al. 2002). Filled dots refer to the observation
for G1, and M15, and  to the hypothetical black hole
in NGC 6752. The horizontal error bar for the black hole in NGC 6752
indicates the uncertainties in the measure of
the central stellar dispersion velocity.
The upper horizontal line refers 
to the upper limit on under-luminous matter  present in the core  
as inferred from the pulsar acceleration measurements. The lower
horizontal 
line is the limit on $\MBH$ imposed by the absence of a rise
of the stellar density expected inside the sphere of influence of the
black  hole (Ferraro et al. 2003a).
}
}
}
\vskip 0.2truecm

As regard to the case of the single intermediate-mass black hole $\MBH$ in NGC 6752,  
we impose a mass of $\sim 500 \msun$.  This value is derived according to 
the relation $\log (\MBH/\msun)=\alpha +\beta \log(\sigma/200 \kms)$
with $\alpha=8.13\pm 0.06$ and $\beta=4.02\pm 0.32$ (Tremaine et al. 2002)
extrapolated down to  $\sigma\approx 10 \kms$.
The central value of the line-of-sight dispersion velocity $\sigma$ 
in NGC 6752 has been estimated to be 
between $2.1-9.7\kms$  (Dubath, Meylan, \& Mayor 1997). 
Recent Fabry-Perot spectroscopy of single stars in NGC 6752 has shown a
flat profile with typical dispersion of $\sim 7\kms$ within the central
$1'$ (Xie et al. 2002), while  proper motion measurements of
stars in the central part of the cluster suggest a much higher value
of $\sim 9-15\kms$ (Drukier et al.  2003).  
In Figure~\ref{fig:fig2} we have drawn 
the $\MBH$ versus $\sigma$ relation and indicated
the upper limit on the central under-luminous mass 
implied by  the pulsar acceleration measurements, together with 
the upper 
limit on the black hole mass $\MBH\simless 1,000 \msun$ 
imposed by the lack of a rise in the stellar density 
inside the core resolved down to a scale
of 0.08 pc (Ferraro et al. 2003a).

\section{Binary-Binary Encounters}
\subsection {The binary pulsar}

Here, we study  the 4-body dynamics of two 
binaries interacting    
under Newtonian gravity.  
Four cases are explored resulting from the combination of 
``projectiles'' occurring in two flavors, and ``targets'' of two types.
The projectile is labelled as \psra  
when it coincides with the observed binary pulsar (in its post-recycling
phase). 
It is labelled  as  \ns when the neutron star, not recycled
yet, is orbiting around a more massive main sequence star
(in its pre-recycling phase).

The binary pulsar \psra  is described as in D'Amico et
al. (2002).
PSR-A has likely experienced a phase of recycling and
of orbital circularization (Bhattacharya \& van den 
Heuvel 1991) that has driven the neutron star to spin
at the observed period of 3.27 ms (CPG).   
PSR-A orbits around a companion star   
of mass $m_{\rm co}=0.2~{\rm M_\odot}$ (a value derived adopting a  
mass of 1.4$\msun$ for the pulsar and an orbit inclination in the
range of 60-90 degrees).  
The binary separation and eccentricity are  $a_{\rm PSR-A}=0.0223$ AU
($P_{\rm{orb,PSR-A}} = 0.86$ days) $e_{\rm
PSR-A}\leq{} 10^{-5},$  respectively (D'Amico et al. 2002).

In the pre-recycling phase, the neutron star (NS) has a mass of $1.4\msun$
and the companion a mass $m_{\rm co}$ of $0.8\msun$ 
consistent with evolutionary scenarios of low-mass binary pulsars
(Tauris \& Savonije 1999).
The initial semi-major axis of the binary [NS,CO] is of $0.03$ AU
corresponding to an orbital period $P_{\rm orb,B}=1.3$ days, below the
bifurcation point (CPG).  The initial eccentricity is $0.7.$ 
Here on, the total mass of the binary will be denoted as
$m_{\rm B}$ (1.6 $\msun$ for \psra, 2.2 for \ns).

\subsection {The \bhbh or \bhcusp binary as target}

The target 
is  either a binary composed of two black holes [BH,BH] (in one case a black
hole-neutron star binary \bhns), or an intermediate-mass black hole
with a  star [BH,*]  orbiting around.

In the first hypothesis,
the initial semi-major axis $a_\bh$ of all our black hole  
binaries \bhbh is of 1 AU, 
unless specified otherwise, and the initial eccentricity is $e_{\bh}=0.7.$
The black hole binary is always  sufficiently
hard 
to provide the right recoil speed $V_{\rm p,PSR-A}$
to 
[PSR-A,CO] (or [NS,CO]). Since  this speed is or the order of 
$30-40\kms,$ the separation  $a_{\bh}$ fulfills the inequality: 
$a_{\bh}\simless 6 \,\xi_\bh (\mu_{\bh}/10\msun)(40\kms /V_{\rm p,PSR-A})^2\,\rm AU$ 
(see eq. [3] and CPG for details; $\mu_{\bh}$ is the reduced mass of the black hole binary).  The binary 
is also wide enough to have a 
coalescence time $\tau _{\rm GW}$ longer
than $\sim 10$ Gyrs which implies 
$a_\bh\simgreat 0.4 \left [(\mu_\bh/10\msun)
(M_\T/100\msun)^2(\tau_{\rm GW}/10 \rm{ Gyr})\right ]^{1/4}\,\rm {AU}$, for an eccentricity
of 0.7. 
This is a necessary condition  since    
the characteristic time-scale for dynamical friction $\tau_{\rm DF}$
to drive \psra
from the current position to the core 
has been estimated $\sim 1 $ Gyr (see CPG; Sigurdsson 2003). This time-scale
imposes 
a lifetime for the black hole binary comparable to the age 
of the cluster itself, unless the binary formed recently.
As an example, for a target 
binary with ($50\msun,10\msun$),
and $e_\bh=0.7$, the suitable interval for $a_\bh$ is (0.3,6) AU.

In the second hypothesis, 
the  target is the central intermediate-mass black hole
of ${\cal{M}}_{\rm BH}\sim 500~{\rm M_\odot}$ 
surrounded by a swarm of bound stars belonging to the {\it cusp}, the 
region 
of influence of the black hole
extending up to a distance $r_{\rm BH}\approx G{\cal M}_{\rm
BH}/\sigma^2\sim 0.02 {\cal M}_{\rm BH,500}/\sigma^2_{10}$ pc,
where $\sigma_{10}$ is the stellar central line-of-sight velocity dispersion in units of 10 km
s$^{-1}.$ 
(Note that even in the absence of a stellar cusp,
a large central black hole $\MBH$ can easily capture a star or 
exchange with a binary star in the cluster 
core, likely a neutron star or a stellar-mass black hole.)
We selected a cusp star (or companion star) of $1 {\rm {\rm M_\odot}}$
tightly bound to the large black hole, moving on a Keplerian orbit with 
semi-major axis $a_{\bh,*}$ of 1 AU, 
and eccentricity 0.7 ($P_{\rm orb, BH}=16$ days). 
The cusp star 
(CS hereafter) is well inside the so called critical
radius $r_{\rm crit}$ where the flow of bound stars percolating across
the loss cone peaks. At distances $r<r_{\rm crit}$  relaxation 
leads to relatively small changes in the integrals of
motion of the stars over a orbital period  (Shapiro \&
Lightman 1977), and  avoid prompt capture by the hole, the loss
cone remaining empty. Thus,  conditions $a_{\bh,*}<r_{\rm crit}$
and $e\simless 0.7$  for a CS 
guarantees  stability of the orbit    
during the characteristic time of the dynamical interaction  $\tau_{\rm enc}\sim
{\sqrt{a_{\bh,*}b}}/\sigma$ with \psra (or \ns), 
where $b$ is the impact parameter of the encounter.
In general, the following
inequalities hold: $P_{\rm{orb,PSR-A}}< P_{\rm
orb,BH}\simless \tau_{\rm enc}<\tau_{\rm rel}$ where $\tau_{\rm rel}$ is the
relaxation time inside the cusp.

\subsection {End-states}

Sampling of the initial conditions is carried on
using the prescriptions outlined in Hut \& Bahcall (1983), and Sigurdsson 
\& Phinney (1993). 
{ 
\vskip 0.2truecm \epsfxsize=8.truecm \epsfysize=8.truecm
\epsfbox{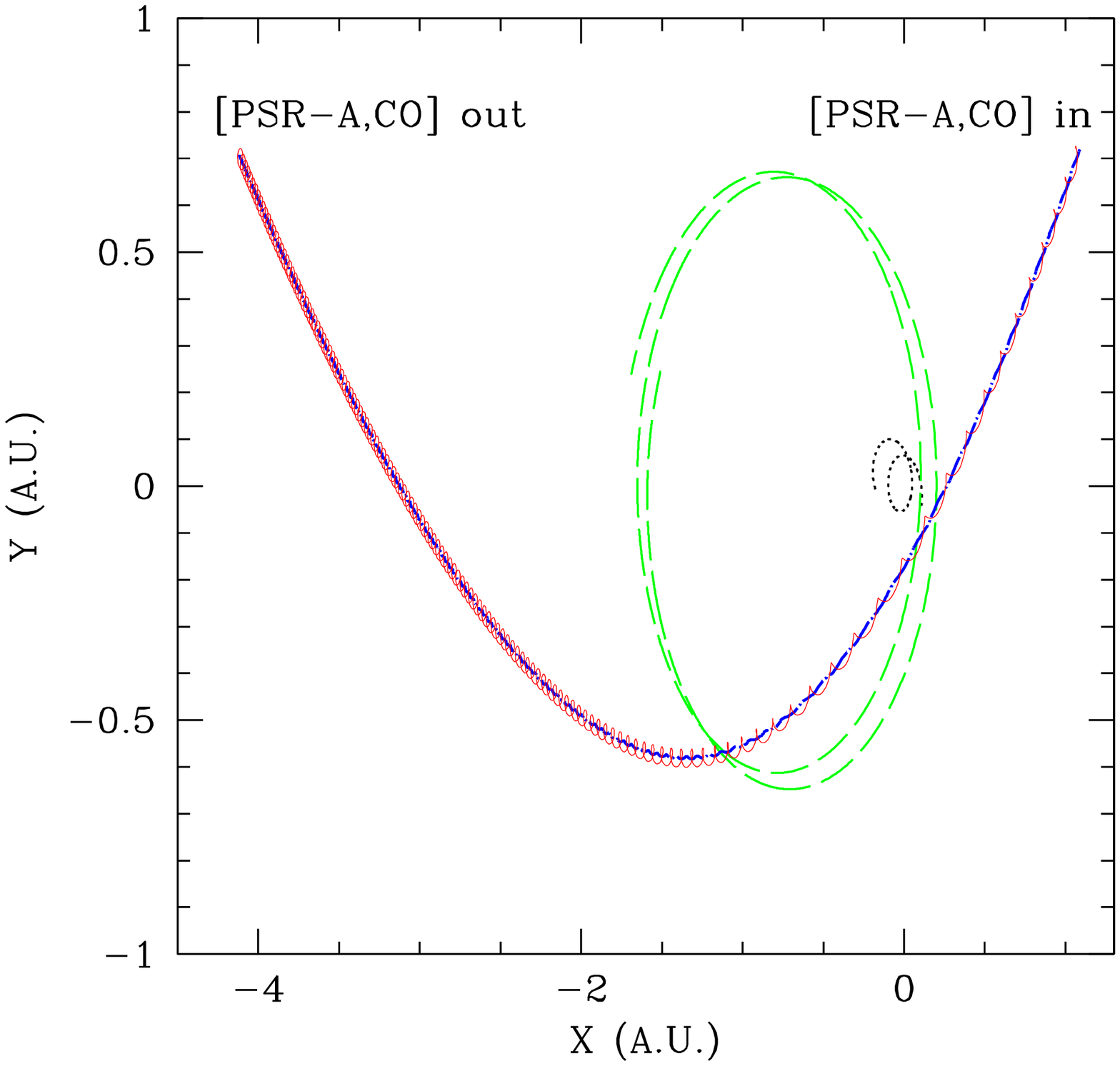}
\figcaption[figure3.eps] 
{
\label{fig:fig3}
\footnotesize 
{
\psra is impinging onto the \bhbh binary of ($50 \msun,10\msun$)
and acquires the thrust to reach the halo of NGC 6752.
}
}
}
\vskip 0.2truecm

The integration method, 
a fourth order Runge Kutta scheme  with adaptive
step size and quality control, maintains the accuracy on total energy
and angular momentum  conservation 
up to  $\Delta E/E_0\sim 10^{-9}-10^{-8}$ and 
$\Delta J/J_0\sim 10^{-11}-10^{-10}$, respectively.  
After a close dynamical interaction, typical values of 
the fractional change of the binding energy of the black hole binary 
$(\Delta E_\bh/E_{0,\bh})$ are of the order of 
$\sim 10^{-2}-10^{-3}$ (here $E_{0,\bh}=G\Mbh\mbh/2a_\bh$). 
During integration, a 4-body to a 2-body 
switch 
is performed in the asymptotic regions, both for the incoming and outgoing 
states. 
The code has been tested  against Sigurdsson \& Phinney  (1993)  
in the case of 3-body scattering.  A trial 
4-body experiment for a reference case
(the $50\msun,10 \msun$ binary) 
has been compared with the output of  STARLAB
\footnote{\url{http://www.ids.ias.edu/~starlab/}}
thanks to the cooperation of Alessia Gualandris and Simon
Portegies Zwart. We found agreement between the
two outputs within the statistical uncertainties
involved in the simulations \footnote{
Alessia Gualandris and Simon Portegies Zwart kinly
accepted to run a comparison model. A detailed 
comparison between the outcomes of the two codes 
will be presented elsewhere.}.

Impact parameters and relative velocities of the two 
interacting binaries have benn  selected 
to ensure an effective energy exchange between the
two binaries leading to post-encounter states 
that can be classified into five main groups: pure
fliesby (FBs), ionizations (IONs), quartets (Qs),
unresolved encounters (UEs) and exchanges (EXs) (the last two
being very rare).
Ionizations can further be sampled in (i) 
resonant ionizations in which [PSR-A,CO] is dissociated 
(their relative separation $\simgreat 20 a_{\rm PSR-A}$),
but  the two stars are bound to \bhbh (or \bhcusp) 
(within a separation relative to the center of mass of black hole binary 
smaller than $30 a_{\bh}$); 
(ii) ionizations with the escape  of the two stars; 
(iii) ionizations with formation of a hierarchical triple  
in which PSR-A (or CO) escapes to infinity leaving CO (or PSR-A) 
bound to the \bhbh (or \bhcusp)
binary. These  triplets may end either in stable or unstable
states (Hut 1993; Mardling \& Aarseth 2001; we return to these systems in $\S 6$). 
Quartets (Qs) correspond to  cases  
where the total energy relative to the centers-of-mass of the binaries is
negative \footnote {Qs can be stable (Hut 1993) 
but are likely to be  perturbed by flying by stars in the dense star
cluster (Bacon et al. 1996).}. 
UEs are very tight Qs occurring when the 
separation of [PSR-A,CO] relative to the black hole(s)
never exceeds $30 a_{\bh}$ after $10^8$ time steps. 
Integration is aborted when   
Qs or UEs appear. On the contrary
IONs and FBs are integrated until the outgoing star(s)
reach the asymptotic state. 
As an illustration, Figure~\ref{fig:fig3} shows a 
FB for [PSR-A,CO] of interest to  our studies.

All end-states, i.e., FBs, IONs, UEs and Qs, 
are recorded to calculate 
relative probabilities, and for the case of FBs we produce
post-encounter distributions for the recoil velocity, 
and eccentricity  of \psra (or \ns). For [BH,BH] the distributions include
IONs, UEs and Qs.  All distributions are normalized to unity.

\subsection{Energy exchange and recoil velocities}

Similarly to the case of 3-body encounters (Hills 1983; Quinlan 1996), 
and given the large mismatch between the  mass 
$m_{\B}$ of the binary pulsar
[PSR-A,CO] (or \ns) 
and $M_{\rm T}$ of [BH,BH] (or [BH,*])
we are led to quantify 
the mean energy exchange per
scattering  of the target binary ([BH,BH] or [BH,*]) as
\be
{\Delta E_{\bh}\over E_{0,\bh}}=\xi_{\bh} {m_{\rm B}\over M_{\rm T}}, 
\ee
where $M_{\rm T}=\Mbh +\mbh$ or $=\MBH+m_*$ for the two cases, 
and   $\xi_\bh$ is derived from our 4-body simulations.
When the black hole binary hardens, most of its energy change will go into
kinetic energy of the light projectile, giving a post-encounter
recoil speed\footnote {To avoid further labeling we will denote with
$V_{\psa}$ the post-encounter velocity of the center of mass of the
binary in the two cases \psra, and \ns.}
\be
V^2_{\psa}\sim \xi_{\bh} {G\mu_\bh\over a_{\bh}}
\ee 
where again $\mu_\bh$ is the reduced mass of either  [BH,BH]
or 
[BH,*].
The cross section, enhanced by gravitational focusing,  is defined as
\be
\Sigma_{\bh}=\pi b^2_{\infty,\rm {max}}\equiv {\tilde {\Sigma}}
\pi (a^2_\bh+a^2_{\rm B}){V^2_{\rm ion}\over V^2_{\infty}}\approx
 {\tilde {\Sigma}}\pi a_\bh {G M_{\rm T}\over V^2_{\infty}}.
\ee
where $b_{\infty,\rm{max}}$  is the maximum impact parameter for which
the fractional energy exchange per scattering $ {\Delta E_{\bh}/E_{0,\bh}} \simgreat 10^{-5};$
$b_{\infty,\rm {max}}$ is estimated at start from  
the analytical expression of Sigurdsson \& Phinney (1993), and later 
is determined more accurately from the numerical scattering experiments.
Here, $V_{\infty}$ denotes
the pre-encounter 
relative velocity between  the two centers of mass of the  binaries; it
is taken close to 
$7 - 10 \kms,$ while 
$V^2_{\rm ion}=[2(m_{\rm B}+M_{\rm T})(E_{0,\bh}+E_{0,\rm B})/(m_{\rm
B}M_{\rm T})]$ is the velocity necessary to dissolve  the  
4-body system (Bacon
et al. 1996).

\section {Results}

\subsection {Post-recycling binary-binary encounters}

{ 
\vskip 0.2truecm \epsfxsize=8.truecm \epsfysize=8.truecm
\epsfbox{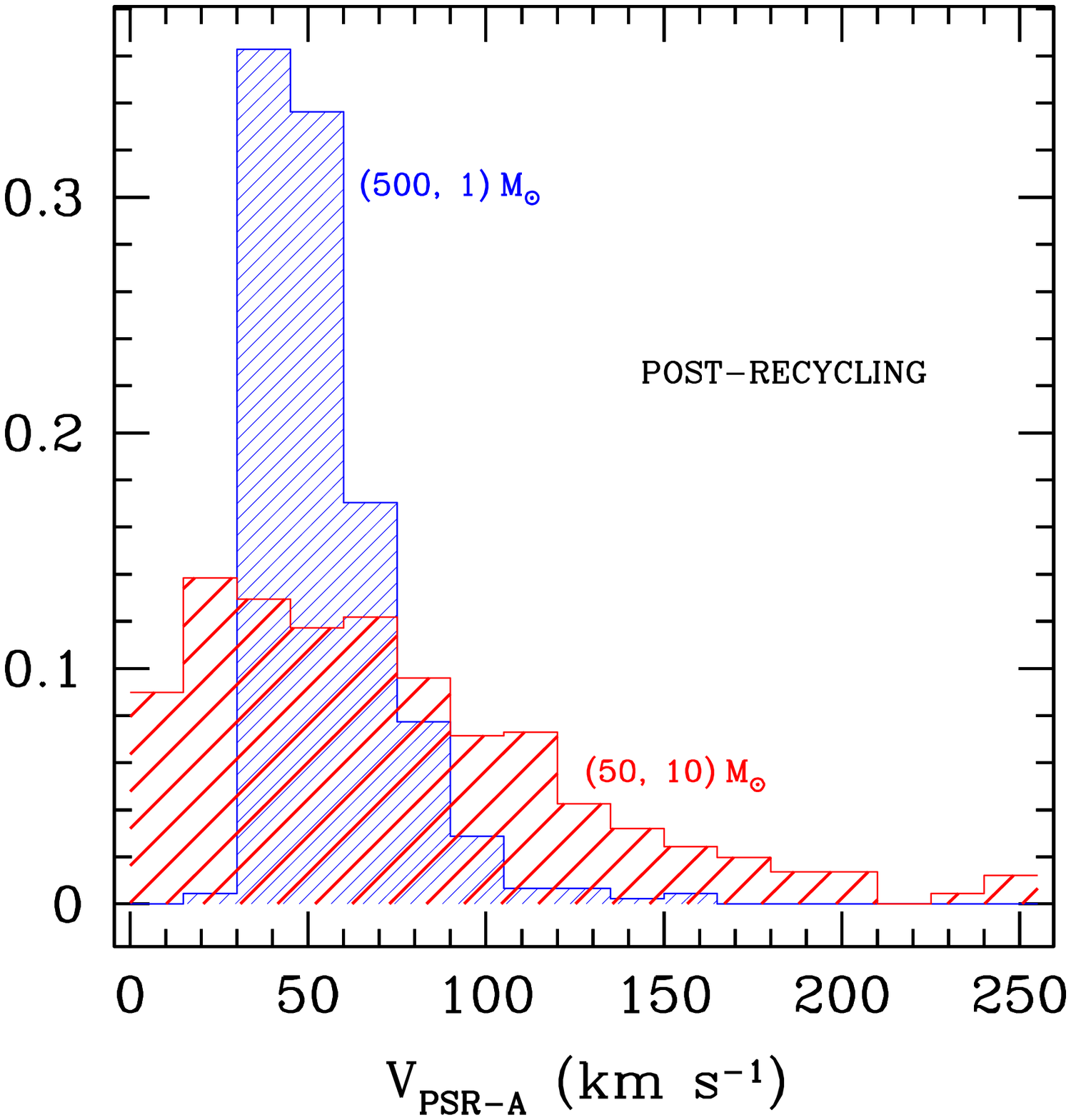}
\figcaption[figure4.eps] 
{
\label{fig:fig4}
\footnotesize 
{
Post-encounter recoil velocity
distribution of [PSR-A,CO] 
including only FBs. \psra scatters off the (50 $\msun$,10 $\msun$)
black hole binary, and off [BH,*]. 
}
}
}
\vskip 0.2truecm

As an example, we first compare 
the end-states between [PSR-A,CO] 
and the stellar-mass black hole binary [BH,BH] of 
($50\msun,10\msun$) with those
resulting from the interaction of [PSR-A,CO] off the black hole
plus  cusp star system [BH,*] of
($500\msun$,$1\msun$).
We recall that the initial 
semi-major axis of the  binaries hosting the  black holes (or the hole and the
cusp star)  
is of 1 AU corresponding to a binding energy of $4.5\times 10^{48}$ erg
(note that \bhbh and \bhcusp  have equal energy in this case).
[PSR-A,CO] at the observed separation of
0.0223 AU has a binding energy of $10^{47}$ erg.
Figure~\ref{fig:fig4} shows the post-encounter distribution of $V_{\psa}$
obtained, for the two cases, collecting only
FBs (the relevant end-states for
the description of \psra).

The distributions  peak at 
$V_{\rm {p},\psa}\sim  30\kms$ (for \bhbh) and $45 \kms,$ 
(for \bhcusp) just around the value of the recoil speed 
necessary to propel the binary pulsar 
in the halo NGC 6752.
The distributions are remarkably asymmetric and we quantified
their dispersion around the peak value in Table~\ref{tbl-1}.
Table~\ref{tbl-1} collects 
post-encounter physical quantities,  while
Table~\ref{tbl-2} collects data on relative 
occurrence probabilities for the  
different end-states.
In general, FBs are associated  
to  less-vigorous/weak-recoil  dynamical encounters compared to IONs.
For \bhbh,
FBs amount to 66\% of all events, while only 11\% of the outcomes
end in IONs (see Table~\ref{tbl-2}).
For \bhcusp, IONs amount to
27\% of the events 
against the 45\%  of FBs.
The
single massive black hole, having 
a stronger tidal field and  a stronger gravitational focusing
(the overall field being closer to a monopole)
imposes a higher
frequency of IONs ending in a narrower velocity distribution
for the FBs.

The post-encounter eccentricity of \psra is remarkably different
in the two cases.
Figure~\ref{fig:fig5} shows the two distributions. In the
case of interaction of \psra with a \bhbh binary, the 
end-state eccentricities peak around  $\sim 2\cdot 10^{-5},$ 
and so \psra preserves the very low initial eccentricity.
The interactions off \bhcusp are instead quite damaging, yielding a 
significant change in $e_{\psa}.$ 
The difference 
can be ascribed, again, to 
gravitational focusing. In fact, in the interaction of \psra off \bhbh
gravitational stresses onto the
incoming binary pulsar are reduced (due to the smaller total mass of the
system) and softened (due to the dipolar nature of
the interaction with the two
BHs), thus causing less damage to the internal orbits of \psra.

{ 
\vskip 0.2truecm \epsfxsize=8.truecm \epsfysize=8.truecm
\epsfbox{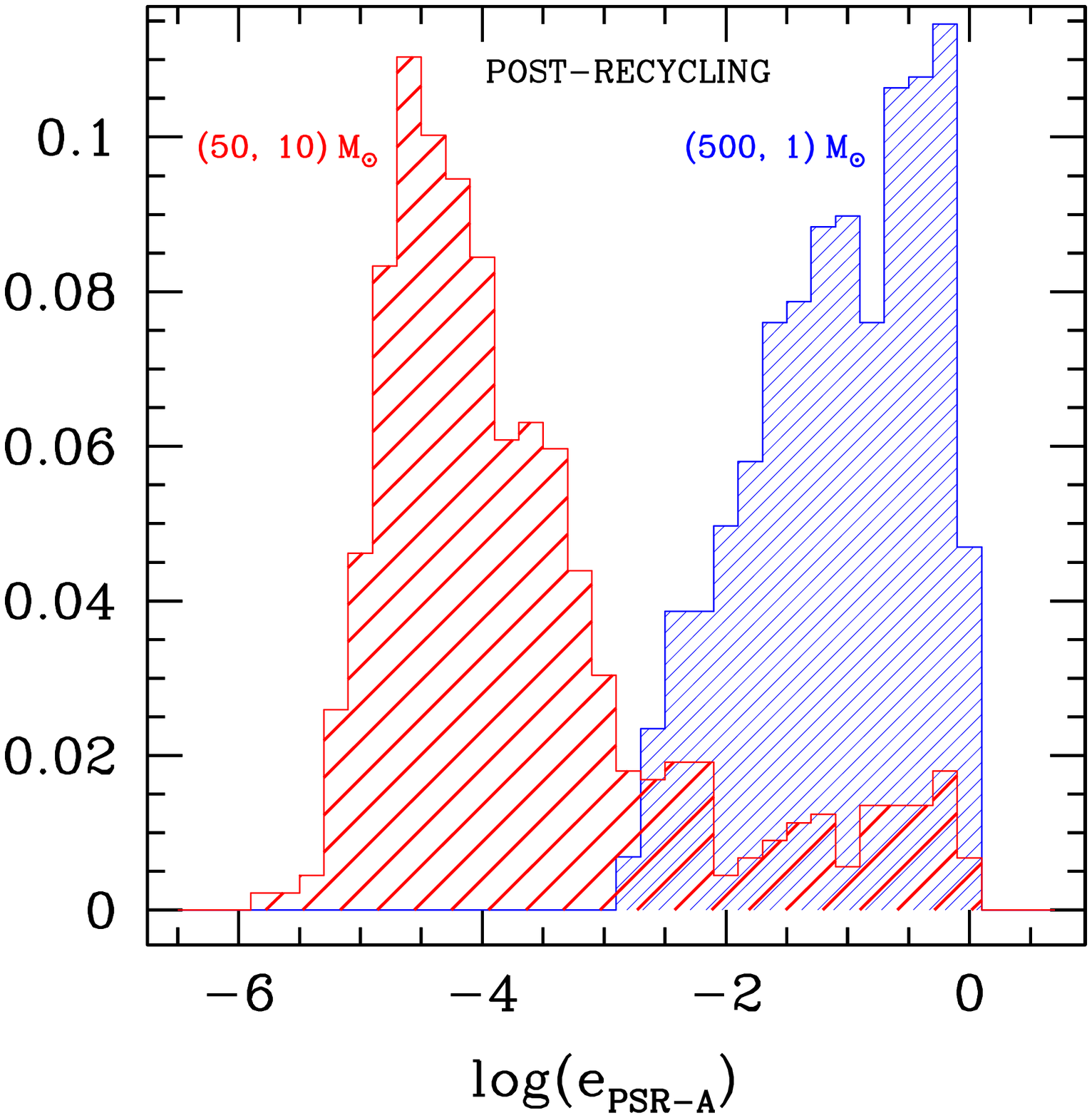}
\figcaption[figure5.eps] 
{
\label{fig:fig5}
\footnotesize 
{
Post-encounter 
distribution of the eccentricity $e$ of  [PSR-A,CO] 
including  FBs only. \psra scatters off [BH,BH] of (50 $\msun$,10 $\msun$)
and off [BH,*].
}
}
}
\vskip 0.2truecm

Table~\ref{tbl-1} surveys results from a series of run carried on 
considering  stellar-mass and intermediate-mass binary black holes.
The recoil speed of \psra increases with increasing $\mu_{\bh}.$ 
But for the most massive binary, i.e., the ($200\msun,10\msun$) case, 
the rise of $V_{\rm p,PSR-A}$ is faster than that implied
by the linear scaling with $\mu_{\bh}$ (eq. [2]) 
suggesting a dependence of $\xi_\bh$ on the total mass of the binary.
For the ($200\msun,10\msun$) binary, 
we find that the peak value of the recoil speed of \psra largely 
exceeds the escape speed. It varies with $a_\bh$ as
\be 
V_{\rm {p},\psa} = V_{\rm {p},\psa}(\rm {1~ AU})
\left ({1 ~{\rm AU}\over a_{\bh}}\right )^{0.44},
\ee
so we can reduce  the magnitude of the recoil
considering FBs off a much wider \bhbh; at a separation $a_\bh$ of 7 AU,
\psra would receive the right pull. Note that 
this intermediate-mass black hole binary does not damage significantly
the eccentricity of \psra despite the fact that the mass of the heaviest 
hole ($200\msun$) is not so far from the mass $\MBH$ of the 
single hypothetical black
hole.

The close 
encounters 
with the binary pulsar exert random impulses
on the \bhbh  binary. 
Linear momentum conservation then imposes a recoil 
velocity to the center-of-mass of the 
\bhbh binary.
Figure~\ref{fig:fig6} shows the skewed distribution of
the recoil velocity for the ($50\msun, \,10\msun$) black hole binary
scattering off the recycled \psra.
Table~\ref{tbl-1} collects characteristic values 
of the recoil speed   
computed averaging over all FBs and IONs.
The rms recoil velocity of the binary black holes
is clearly in excess to its equipartition value 
(as expected for a point mass in a star background).    
It will be moderated by frictional drag and we defer to $\S 7$ for a discussion  
(see Merritt 2002 for a study  of the Brownian motion
of a massive binary near equipartition).

As described  by equation (2), the target black hole binary 
transfers its gravitational binding energy to the binary pulsar (mainly in 
the form of kinetic energy
of the center-of-mass of the projectile binary).
In the last row of Table~\ref{tbl-1}  we collect the values of
$\xi_\bh$ obtained averaging over all 
encounters, and in bracket the values obtained 
selecting only FBs. The significantly lower value of
$\xi_{\bh}$ when inclusive of all events is 
related  to the formation of  triple systems: 
The formation of a hierarchical triple requires loss in binding
energy of the target binary that widens to incorporate the third star.

The black hole binary transfers  its internal orbital angular
momentum
$J_{0,\bh}$ to the  orbit of the outgoing 
binary pulsar (that in this context 
can be treated as a point-mass, having negligible
internal angular momentum).  
The initial Keplerian angular momentum of the binary black holes,   $J_{0,\bh}\sim
10^{55}(M_{\rm BH,200})(m_{\rm BH,10})(
M_{\rm T,210})^{-1/2}(a_{\rm BH,1{\rm AU}})^{1/2}\rm {g\, cm^2\, s^{-1}}$
(subscripts refer to units consistent with a [$200\msun,10\msun$] binary)
exceeds by a factor $\sim 10$ 
the orbital angular momentum of the pulsar
$J_{\rm 0,B}\sim 10^{54}m_{\rm B,1.6}(b_\infty/30\, 
{\rm AU})(V_{\infty}/
10\kms)\rm {g\, cm^2 \,s^{-1}},$  where the impact parameter 
$b_\infty\sim (0.5a_\bh GM_\T/V^2_\infty)^{1/2}$ and $V_\infty$ define the initial
unperturbed hyperbolic orbit of the projectile.
We find that the fractional  angular momentum change of the black hole 
binary $\Delta J_\bh/J_{0,\bh}$ amounts to  
$\sim 0.01-0.1$ (see Table 1).  This is comparable or
less than $J_{\rm 0,B}/J_{0,\bh}$ so the binary pulsar can only marginally 
modify  its post-encounter  orbital angular momentum vector.  A more massive binary black hole
(with mass $\simgreat  1000 \, \msun$) would be necessary 
to produce  a sizeable  
change in the angular momentum vector of the binary pulsar. 
Stars of mass near turn-off
($\langle m \rangle \sim 0.5 \msun$) could instead be affected 
(see Miller 2003 for a discussion on the possible origin of rotation observed 
in the core of globular clusters and attributed to a binary black hole).

{ 
\vskip 0.2truecm \epsfxsize=8.truecm \epsfysize=8.truecm
\epsfbox{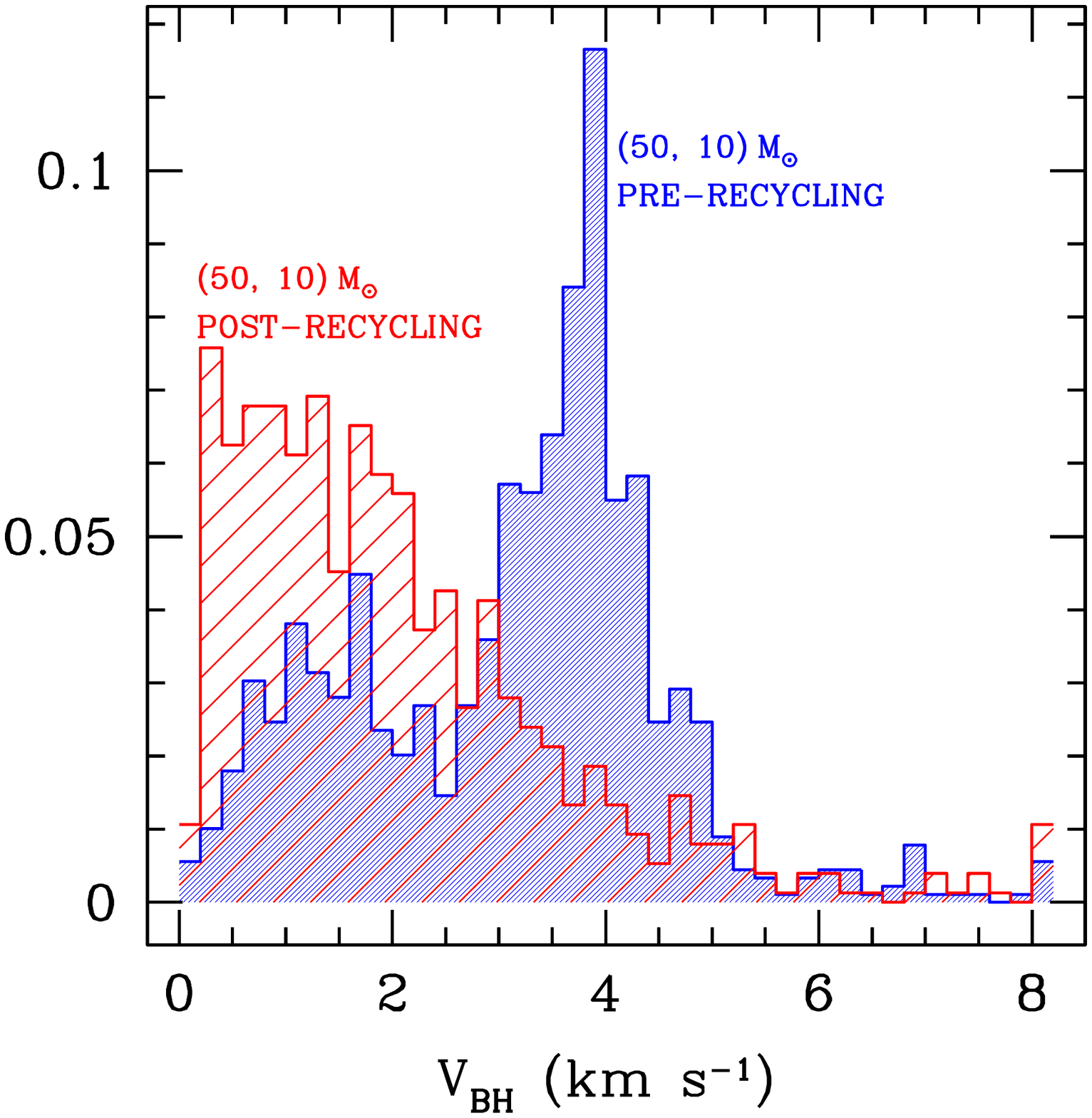}
\figcaption[figure6.eps] 
{
\label{fig:fig6}
\footnotesize 
{
The post-encounter recoil velocity of the \bhbh binary 
(50 $\msun$,10 $\msun$) in the pre- and post-recycling scenario.
}
}
}
\vskip 0.2truecm


\subsection {Pre-recycling binary-binary encounters}

In this section we explore  binary-binary encounters
before the neutron star is recycled, studying two cases: 
a ($50\msun,10\msun$) binary, and the single black hole \bhcusp.

As shown in Table~\ref{tbl-3} 
the thrust received by \ns 
is within the correct range for expulsion of the binary
(progenitor of \psa) in the halo of NGC 6752, in either cases 
(\bhbh or \bhcusp).  Despite the large difference in mass between
$\MBH$ and the mass $\Mbh$  of the heavier black hole
in the binary, the peak velocities
$V_{\rm p,PSR-A}$ are similar since recoil
is sensitive to the reduced mass of the system.
 
FBs off the \bhcusp binary lead to
a significant increase of the post-encounter eccentricity 
and to a change in the binding energy of \ns.
We have noticed that 20\% of these FBs produce a \ns binary
with orbital separation such that the companion star fills  its
Roche lobe.
This hints in favor of a ``collisionally induced'' recycling 
process, as suggested by Sigurdsson (2003).
(Note that recycling lasts for a time less or  
comparable to the lifetime 
of the binary pulsar in the cluster halo according to dynamical
friction, making this picture consistent.)
Orbital changes of \ns are much less significant for the \bhbh case, and 
``induced'' recycling has a frequency of less than 10\%.

In the pre-recycling case, 
the higher frequency of IONs found  
causes a remarkable increase
in the recoil velocity of the \bhbh binary
that can reach peak values $\sim 4 \kms$ (Table~\ref{tbl-3}).
As illustrated in Figure~\ref{fig:fig6} for the ($50\msun,10\msun$) binary,
the distribution of the recoil velocity  
displays two humps, the first associated to FBs,
and the second to IONs.
IONs can thus imprint significant  
non-equilibrium velocities on the black hole
binary.

\section {The statistics of the encounters }

In the post-recycling scenario, FBs are statistically more frequent. 
In general, with increasing mass of the black hole binary and 
decreasing mass ratio $\mbh/\Mbh,$ 
the frequency of IONs increases; from 8\%-12\%
if the target is a \bhbh binary, up to   $\sim 30\%$ 
in the \bhcusp case.
Triple systems have a high probability of formation, among IONs,
and they will be described in $\S 6$. 
Qs account for $\sim 20-30\%$ of the end-states; 
many of these systems are wide and loose so that the interaction
with the gravitational potential of the cluster causes their
fission into two separate binaries.
In the pre-recycling case, the statistics reverse, 
causing IONs to become more important than FBs with a frequency as
high as $85\%,$ for the \bhcusp case. 
FBs comprise only 7\% of the events in the black hole-cusp star hypothesis,
and 23\% in the \bhbh case.

Having computed  the frequency of FBs relative to all possible end-states,  
we can estimate the collision rate 
of the binary off the \bhbh or \bhcusp 

\be
{\cal {R}}_{\rm coll}={1\over \tau _{\rm coll}}
\sim {\tilde {\Sigma}}_{\rm BH, FBs} 2  \pi a_{\bh} GM_{\rm T}{
f \, n_{\rm NS}\over V_{\infty}},
\ee 
where  $n_{\rm NS}$ 
is the number density of neutron stars
impinging onto the target black hole(s) and $f$ the fraction in
binaries with a star.
The density of neutron stars 
is largely unknown given
the uncertainties in their retention fraction at birth,
but its value is expected to be larger than the density $n$ of stars (Sigurdsson
\& Hernquist 1993). The value of $f$ is even more uncertain.
To circumvent the problem of estimating $n_{\rm NS}$ and $f$ 
we list, in Table~\ref{tbl-5}, the minimum density 
of recycled neutron stars $f\, n_{\rm NS}$ necessary 
to have an encounter probability
comparable to that of detecting \psra 
at the current position. Since the estimated time of
dynamical friction $\tau _{\rm DF}$ is $\sim 1 $Gyr, this minimum
density is computed imposing a rate ${\cal R}_{\rm coll}\sim $Gyr$^{-1}$ 
and a black hole binary separation $a_{\bh}$  equal to 
the maximum  necessary to acquire 
a recoil velocity 
of the order of $\sim 35\kms.$
These minimum  densities cluster around  $50-10^3$ pc$^{-3},$
implying a ratio $f n_{\rm NS}/n$  
which is less than that expected from dynamical arguments 
(Sigurdsson \& Hernquist 1993).
Note that  in the case of  pre-recycling, $f n_{\rm NS}$ 
can be much larger, since there are no constraints on 
the spin history of the neutron star.

\section {Exotic end-states}

A black hole binary in a cluster is a catalyst for the
formation of exotic triple systems composed of the two black holes and a star. This is a
consequence of the high frequency 
of ionization events in binary-binary encounters  
ending with the capture of one of the two 
binary stars.
Table 2 and Table 4 
give in bracket the fraction of triple systems that form over the total.
These comprise  from $\simless 5\%$  up to $30\%$ of all events
in \bhbh binaries.
The question is now whether the triplet that forms 
is long-term stable, or short-term stable, i.e., 
destined to lose one of its components after a few outer-orbital times 
(Mardling \& Aarseth  2001).

In our experiments we have focused attention to binary-binary encounters leading  to
the formation of triplets with an active pulsar (in the post-recycling case)
orbiting around one of the two black holes \footnote {Hierarchical triples
composed of two black holes and a main sequence star or a pulsar will be  
studied in detail in a separate paper
(Mapelli, Colpi, \& Possenti 2003).}.  We then extract from the sample
those triplets that fulfill the long-term 
stability criterion 
following Mardling \& Aarseth (2001; their eq. [90]).
The stable triplets are more frequently composed by the millisecond pulsar 
orbiting around the heavier black hole (forming the inner binary) with the
lighter black hole (the outer binary) orbiting  around  the inner pair.
The typical parameters of the inner binaries 
found in our 
simulations  have orbital periods
ranging between  10 to 100 days  corresponding to
orbital separations of 200 to 10,000 light seconds.
The orbital acceleration imparted to the millisecond pulsar by the black hole would 
not hamper its detection by current
deep globular cluster pulsar surveys (Possenti et al. 2003; 
Ransom 2003; Jacoby 2003).  
Figure~\ref{fig:fig7}  shows the formation of a  triplet for 
the ($50 \msun, 10 \msun$) binary. 
Triplets with the millisecond pulsar and the light
black hole as inner binary are long-term stable only if 
their orbit is very tight.
The stronger acceleration suffered by the millisecond pulsar
due to its proximity to the hole 
would induce very rapid changes in the apparent spin period along the orbital motion  
impliying a strong bias against the discovery of such pulsars$^6$.

MSP-black-hole binaries 
have long been searched in globular clusters 
as they can form via 3-body exchange interactions between single black holes
and binary pulsars (Sigurdsson 2003).
We here for-see the possibility of detecting  
a hierarchical triple MSP-black-hole-black-hole system that would provide a direct
measure of the mass of the heaviest black hole. 

{ 
\vskip 0.2truecm \epsfxsize=8.truecm \epsfysize=8.truecm
\epsfbox{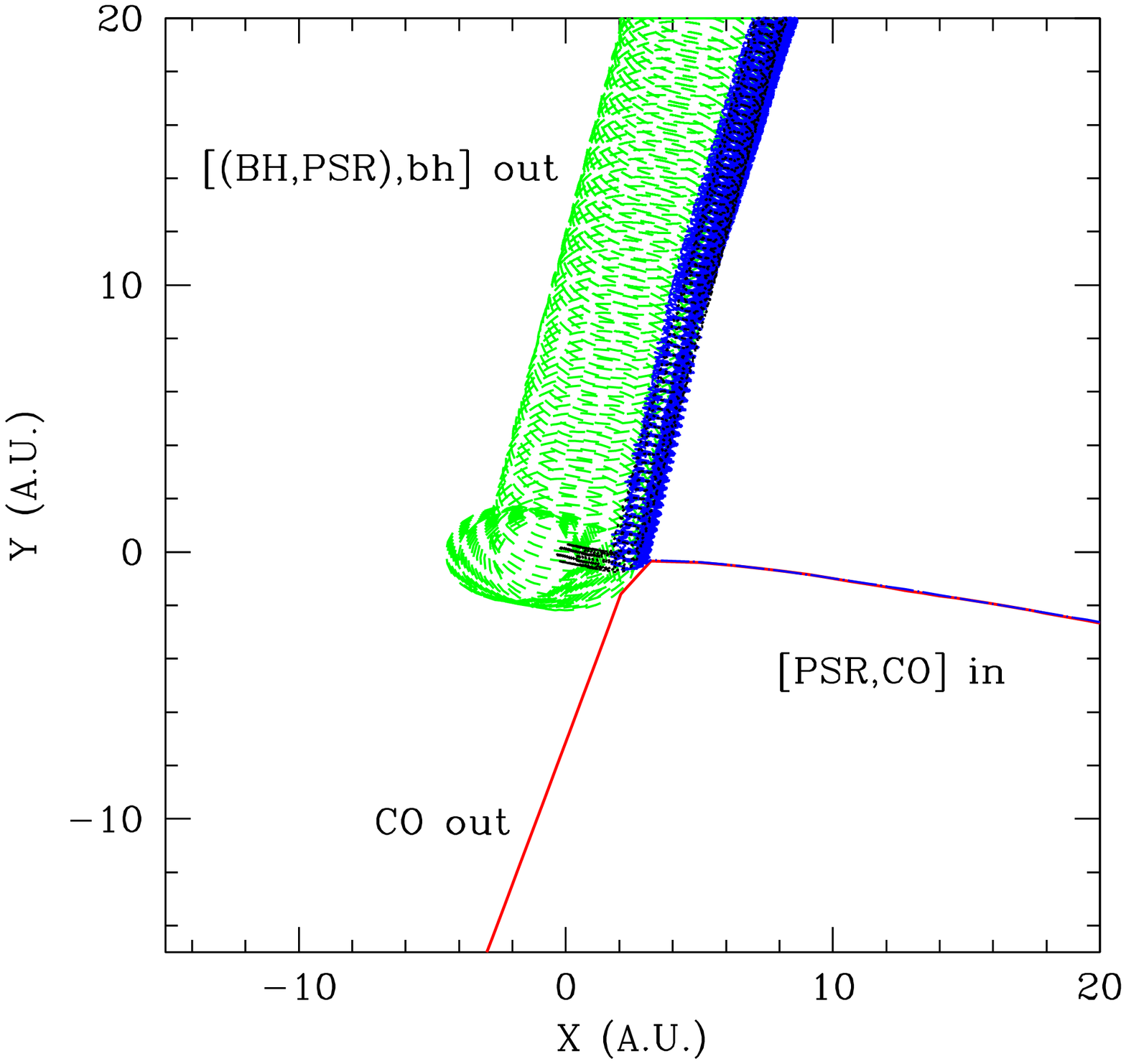}
\figcaption[figure7.eps] 
{
\label{fig:fig7}
\footnotesize 
{
\psra is scattering off the \bhbh binary ($50\msun,10\msun$)
and the ordinary star escapes.
A triplet forms with the millisecond pulsar orbiting around
the heavier black hole.    
The inner binary  has a semi-major axis 
of 0.28 AU and eccentricity 0.97. 
In the outer binary, the  
light black hole orbits around the center-of-mass of the inner
binary with a semi-major axis of 5.1 AU,
a periastron separation of 1.53 AU  and eccentricity 0.7. 
The triplet fulfills the long-term
stability condition of Mardling \& Aarseth (with a ratio
of the periastrion separation of the outer binary to the critical  separation 
given by eq. [90] equal  to 1.2, so greater than unity).
}
}
}
\vskip 0.2truecm


Considering  the case of a single intermediate-mass black hole in the cluster core,
binary-binary encounters of the type studied here can lead to the ``capture'' of 
a millisecond pulsar thanks to the interaction of the binary pulsar with
the start belonging to the cusp. This is a possibility that will be studied in 
more detail, introducing a model for the cusp.

\section {Linking pulsar's accelerations to the 
intermediate-mass  binary black hole  hypothesis}

The single intermediate-mass black hole $\MBH\simless 500$  
can  account  for
a  portion of the unseen matter  required for
explaining the spin derivatives of PSR-B, PSR-E (and perhaps of PSR-D).

Alternatively, one may wonder whether a suitably located perturber
such as a black hole binary  
can accelerate  the two pulsars. (This would reduce the demand of 
a significant amount of under-luminous matter in the core of NGC 6752.)
The perturber should imprint
a line-of-sight  acceleration  $GM_{\rm T}/l^2\sim c\vert {\dot{
P}}/P\vert .$
Considering the value of $\vert \dot P/P\vert=9.6\pm 0.1\times 10^{-17}$ 
for the two pulsars
(PSR-B and PSR-E), and  a separation $l\sim 0.03$ pc (D'Amico et al. 2002), 
comparable to the projected distance
between the two pulsars, we can infer the minimum mass that the \bhbh binary
should have to impart the observed acceleration; 
$M_{\rm T}\simgreat 180\msun$. Thus, the ($200\msun,10\msun$) binary has
the right total mass.
Assuming  an harmonic
potential for the central region of the
cluster with  uniform stellar mass density $\langle m\rangle \,\, n$ of  $\sim 10^5~{\rm
M_\odot~pc^{-3}}$ (where $\langle m\rangle$ is the mean stellar mass), 
the minimum recoil velocity for moving 
the black hole binary  from the
center of the potential well to the pulsar 
projected locations (at $r\sim 0.08$ pc) is 
\be
\langle V^2_{\bh,{\rm {min}}}\rangle ^{1/2}\,\sim
4 \left ({ r\over 0.08 \rm {pc} }\right) \left 
({\langle m \rangle \,\,n\over 10^5\msun~\rm{pc^{-3}}}\right )^{1/2}
\kms. \ee
Table~\ref{tbl-1} and Table~\ref{tbl-3}
show that a \bhbh binary of the required  mass  has typical
$V_{\rm BH}\sim 1 \kms$  ($2\kms$) in the post- (pre-) recycling
scenario, lower than the value required.
However 
the black hole binary is subject 
to repeated close encounters with cluster stars before 
dynamical friction drives it 
toward the center of gravity
of the cluster (Merritt 2002).
This depends on the comparison between
the dynamical friction timescale 
$\tau_{\rm DF}\sim \sigma^3/[1.89G^2 \langle m\rangle  n M_{\rm T}]$ 
and the collision time off  stars  
$\tau_{\rm coll}\sim \sigma/[2\pi GM_{\rm T}n a_{\bh}].$
Their ratio 
\be
{\tau_{\rm DF}\over \tau_{\rm coll}} 
\sim 5 \left ( {a_{\rm {max},\bh}\over 7 \,{\rm AU}}\right )
\left ({\sigma\over 10 \kms}
\right )^2\left({0.5 \msun\over \langle m\rangle }\right)
\ee
is above unity but uncomfortably close.
Because of this non-equilibrium dynamics, an 
intermediate-mass  black hole 
binary of ($200\msun,10\msun$) 
may random walk up 
to the current position of PSR-B and PSR-E.
This scenario is possible but remains fine tuned.

\section{Black hole binary  evolution}

In this Section we address the issue of the survival of our 
stellar and intermediate-mass black hole
binaries  against hardening by stellar collisions.  

\subsection {Binary hardening}

The last black hole binary that remains in the cluster core
(Portegies Zwart \& McMillan 2000) 
has a large cross
section and interacts with the cluster stars.
Thus, we expect it to
harden progressively ejecting stars and emitting gravitational
waves;  at the same time stars may be tidally captured.
{ 
\vskip 0.2truecm \epsfxsize=8.truecm \epsfysize=8.truecm
\epsfbox{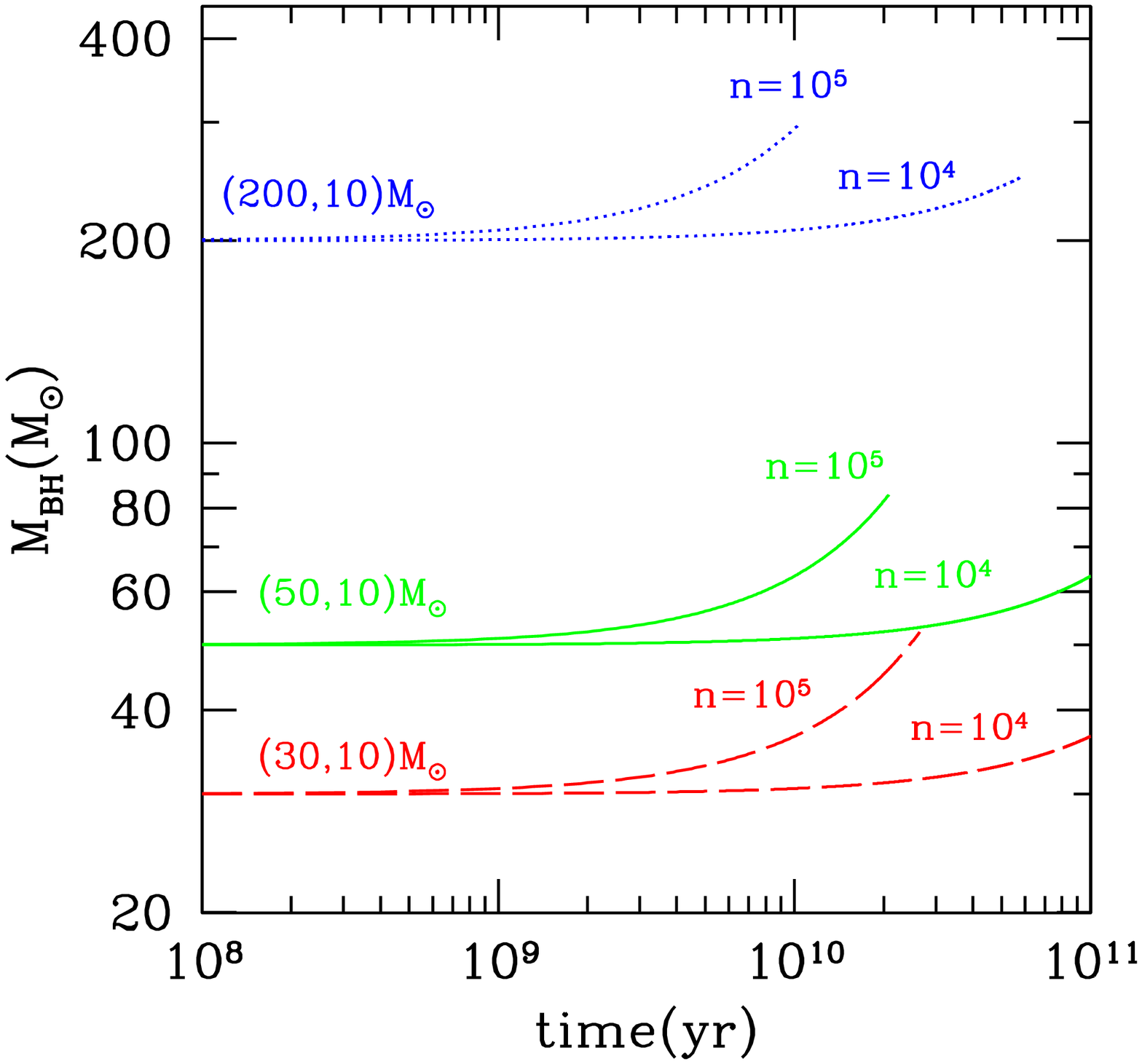}
\figcaption[figure8.eps] 
{
\label{fig:fig8}
\footnotesize 
{
Black hole mass growth, 
$\Mbh$ versus time $t$ (years) in a cluster with background stars 
with dispersion velocity $\sigma=7\kms$ and density $n=10^{4}$ and 
$10^5$ pc$^{-3}$. The dispersion velocity of NSs and WDs is obtained, 
in conditions of thermal equilibrium, re-scaling the stellar dispersion 
velocity as $\sigma_i=\sqrt{\langle m \rangle /m_i}\,\sigma$, where $\langle m\rangle $ is the
mean  stellar mass and $i$=NS, WD. All lines terminate at the
time of coalescence of the two holes by emission of GWs.
}
}
}
\vskip 0.2truecm


It is the heaviest black hole in the binary that
preferentially eats (unbound) stars because of
its higher gravitational focusing. 
This leads to a secular change
in the mass and orbital parameters.
We mimicked  evolution, 
solving the equations for 
the binary separation $a_\bh$ and  mass  $\Mbh$ as a function
of time 
\be
{d a_\bh\over dt}=-2\pi\xi_\bh G a_\bh^2\sum_i {m_i n_i \over \sigma_i}-
{64 G^3 \Mbh\mbh M_{\T}\over c^5 a^3(1-e^2)^{7/2}},
\ee

\be
{d \Mbh\over dt}=2\pi G \Mbh^{4/3} \sum_i m_i^{2/3}{n_i r_i \over \sigma_i}.
\ee
where the sum over index $i$ is inclusive of 
all stars, i.e., main sequence stars, white dwarfs and
neutron stars, distributed according to
the Salpeter IMF, and in thermal equilibrium.
The capture rate on $\Mbh$ refers to a cross section equal to $2\pi G\Mbh
r_{i,{\rm t}}/\sigma^2_i$ where $r_{i,{\rm t}}=r_i(\Mbh/m_i)^{1/3}$ is the
tidal radius of the star considered. We assume that the entire star is eaten.
Figure~\ref{fig:fig8}  shows $\Mbh(t)$ for two values of
the stellar density $n$ (treated as a constant in time).
We found that a secular increase in the mass of the binary black hole
becomes appreciable only  if the initial mass 
$\Mbh(0)>20\msun,$ under the
background conditions imposed.
A ($50\msun,10\msun$) or ($30\msun,10\msun$) binary  
avoid coalescence and its separation remains
$\simless 1$ AU for an appreciably long time.
An ``intermediate-mass'' mass black hole binary 
(such as that considered in $\S 7$ with 
$\Mbh\simgreat 200 \msun$ at the start) should 
have already
terminated its life or be on the verge
of coalescing, ending in a single black hole, and, curiously, should have
grown to a mass close to $\MBH.$  
Clearly, binary evolution is rather sensitive to the background density
which is itself evolving with time (a fact that we did not include
in this simplified treatment). Thus, the dynamical evolution of an intermediate-mass
black hole binary should thus be addressed using a full N-body code.
Lighter ``stellar-mass'' black hole binaries are still in the
hardening phase and the mass increase is not significant. If 
our binaries will coalesce by GW emission, they will remain always far from
the critical regime (corresponding to mass ratios $\mbh/\Mbh$ around 0.385)
at which the gravitational rocket effect comes into play (Fitchett \&
Detweiler
1984), and will avoid ejection from the cluster 
due to release of 
non-zero net linear momentum.

\subsection {Star ejection}

Since black hole hardening (occurring at a rate 
$da_\bh/dt=-2\pi\xi_{\bh}G\langle m \rangle na^2/\sigma$) 
is followed by stellar ejection,
a correlation exists between the mass lost in stars 
\be
M_{\rm deficit}\sim {M_{\rm T}\over \xi_{\bh}}
\ln\left ({a_\bh(0)\over a_\bh(t)}\right )
\ee
and the total mass $M_\T$ of the binary black hole (treated as a constant).
In equation (11), $a_\bh(0)$ refers to the maximum semimajor axis at which the interactions
are strong enough to eject stars. 
When  $a_\bh(t)\sim a_{\rm GW},$
$M_{\rm deficit}$ is the mass that needs to be ejected 
in order to drive the binary toward coalescence by GWs.
If NGC 6752 still hosts a  binary black hole,
one may  ask whether the ``deficit'' in stars created by their ejection
from the cluster would  alter appreciably 
the stellar density profile or 
velocity field in the central regions.

If, similarly to what claimed in bright ``core-ellipticals'',
a binary black hole turns an otherwise power-law 
density profile into a core (or shallower
lower-low)
due to the ejection of stars (see Ferrarese et al. 1994;
Milosavljevi\'c \&  Merritt 2001; Milosavljevi\'c, Merritt, Rest, van den
Bosch 2002 for details), then one can verify if 
the `` mass deficit'' (defined
as the mass in stars that would need to be removed from  
an initially power-law profile in order
to produce the observed core) correlates with the mass of the binary
black hole according to equation (11). De-projecting the power-law+core 
profile of Ferraro et al. (2003a; Figure 6) 
we  
infer for NGC 6752 a deficit of  300 ``missing'' stars.
This is close to  the mass that would be
ejected by one of our  binary black holes (eq. [11]). 
Since the central relaxation time is of the order of the ejection time,
a flow of stars can refill an underlying  power-law, but 
gravitational heating (Spitzer 1987) 
induced by  star loss can counteract  this process.
Only  a detailed cluster model  
with a core binary black hole
can address
this question in detail, and 
should be designed specifically for NGC 6752.

\section{Conclusions}

What have we learned from this analysis ?

(a) A stellar-mass binary black hole of ($50\msun,10\msun$) can
imprint the right thrust to PSR-A in a gentle flyby leaving almost
unperturbed its eccentricity.  Fliesby off this binary can occur at an
acceptable rate without imposing unrealistic conditions on the neutron
star density inside the central 0.1 pc. This binary black hole long
lives in the cluster.  Lighter black hole binaries or binaries with a
black hole and a neutron star have interaction rates smaller by a
factor $\sim 10$ and can be excluded (see Table 5).

(b) An intermediate-mass binary black hole of ($200\msun,10\msun$) is
more aggressive on the binary pulsar, as it tends to impart large
recoil speeds to any incoming particle.  Thus, it has to be
sufficiently wide ($a_\bh\sim 7-10\, \rm {AU}$) to propel \psra
(or\ns) in the halo of NGC 6752 at the speed requested. Such binary
demands the lowest neutron star density to allow one ejection every
billion years (Table 5).  The survival of such binary black hole in
the cluster however is uncertain: The binary may have merged into a
single black hole in its interaction with the background stars: its
presence in the cluster implies either recent formation, or hardening
on a time scale comparable to the age of the cluster.  This binary may
become the seed upon which our hypothetical ``single''
intermediate-mass black hole has grown.

(c) A single, intermediate-mass black hole of $\MBH\simgreat 500\msun$
is a possibility.  The recoil speed of PSR-A falls in the correct
interval and considering its weak dependence on $\MBH$ (recoil is more
sensitive to the reduced mass of the system) even a $1000\msun$ would
fit in this picture.  The gravitational encounter with PSR-A must
occur before recycling as the binary hosting the neutron star suffers
a considerable change in its eccentricity.  Circularization can be
achieved later, during the phase of Roche lobe mass transfer: we find
that the binary is left in an outgoing state such to favor
``scattering induced recycling''.

The considerations in (a,b,c) do not address the problem of the
acceleration of PSR-B and PSR-E. Can we provide a self-consistent
picture ?  If there is a stellar-mass binary black hole in the cluster
propelling PSR-A, this binary can not act as a local perturber on the
two pulsars. It is too light.  The ``best'' candidate for the local
perturber hypothesis is our intermediate-mass binary black hole (the
$200\msun,10\msun$ binary). The binary is sufficiently light so it can
still random walk across the core to produce the local acceleration of
the two pulsars.  If instead the acceleration of PSR-E, PSR-E (and
perhaps PSR-D) is caused by the overall effect of the cluster
potential well, the single intermediate-mass black hole plays a role
but some additional mass should be present in the form of collapsed
remnants such as white dwarfs and neutron stars.

Black hole binaries are special {\it catalysts} for the formation of
``stable'' triplets. The interaction of a binary pulsar with an
intermediate-mass binary black hole can create an extraordinary
system: {\it a millisecond pulsar-black hole-black hole hierarchical
triple} (the companion star leaving the system).  The single
intermediate-mass black hole may, on the other hand, host in its cusp
a millisecond pulsar.  In this perspective, timing measurements on
this ``planetary-like-pulsar'', either member of a triplet or of the
cusp, would give the unique possibility of discriminating the nature,
either single or binary, of the black hole(s), providing also a
reliable estimate of their mass.  This opportunity exists in globular
clusters contrary to the case of the Galactic Center, where the signal
from a millisecond pulsar belonging to the cusp of our central black
hole would be completely smeared out by scattering in the interstellar
medium.

A binary black hole may steadily perturb its environment ($\S$ 8.2).
The distinction between a stellar-mass and an intermediate-mass binary
black hole is subtle and difficult to separate out.  Accurate
evolutionary models of clusters hosting a binary black hole of various
masses need to be developed as they can indicate the fingerprints left
by such exotic binaries. They also may provide clues for
discriminating between the single versus binary nature of the central
black hole(s) that may inhabit NGC 6752.

During the revision of this manuscript Ferraro et al. (2003b) reported
the discovery of the optical identification of the companion star to
PSR-A, a helium white dwarf. The derived cooling age ($\sim 1.2-2.8$
Gyr) suggests that the dynamical encounter responsible for the
ejection of PSR-A into the cluster halo occurred after the neutron
star was recycled or was triggered at the time of ejection. In the
last case the binary should have suffered very rapid evolution.

\acknowledgments{The authors wish to thank Cole Miller (the Referee)
and Simon Portegies Zwart for stimulating discussions and a critical
reading of the manuscript, and Steinn Sigurdsson for advise during the
early stages of the work.  They also warmly thank Alessia Gualandris
and Simon Portegies Zwart for having shared data on binary-binary
encounters obtained with STARLAB and made the comparison with our
direct code possible.  We acknowledge financial support from the
Italian Space Agency under the grants I/R/037/01 and I/R/047/02 and
the Italian Minister of Research (MIUR).}


\newpage
\newdimen\minuswidth    
\setbox0=\hbox{$-$}

\newpage

\begin{deluxetable}{lllllll}
\scriptsize 
\tablewidth{0pt}
\tabcolsep 0.12truecm
\tabcolsep 0.12truecm
\tablecaption{End-states in the post-recycling scenario. \label{tbl-1}}
\tablecolumns{7}
\tablehead{
 & \colhead{($500\,{}\textrm{M}_{\odot{}},\,{}1.0\,{}\textrm{M}_{\odot{}}$)}&
\colhead{($50\,{}\textrm{M}_{\odot{}},\,{}1.4\,{}\textrm{M}_{\odot{}}$)} &
\colhead{($30\,{}\textrm{M}_{\odot{}},\,{}3\,{}\textrm{M}_{\odot{}}$)}   & 
\colhead{($10\,{}\textrm{M}_{\odot{}},\,{}10\,{}\textrm{M}_{\odot{}}$)}   &
\colhead{($50\,{}\textrm{M}_{\odot{}},\,{}10\,{}\textrm{M}_{\odot{}}$)} &
\colhead{($200\,{}\textrm{M}_{\odot{}},\,{}10\,{}\textrm{M}_{\odot{}}$)} 
}
\startdata

$^{\rm a}\textrm{V}_{\textrm{p,}\,{}\textrm{PSR-A}}$ ($\textrm{km}\,{}\textrm{s}^{-1}$) & $45^{+30}_{-20}$ & $15^{+30}_{-10}$ & $15^{+40}_{-10}$  & 
$18^{+25}_{-10}$  &  
$30^{+10}_{-20}$ & $90^{+60}_{-60}$  \\

$^{\rm b}\langle{}\textrm{V}_{\textrm{PSR-A}}\rangle{}$ ($\textrm{km}\,{}\textrm{s}^{-1}$) & 
$55^{+29}_{-13}$ & $27^{+44}_{-15}$ & 
$35^{+62}_{-21}$ & 
$32^{+112}_{-16}$ &  
$73^{+78}_{-42}$ & $101^{+116}_{-40}$ \\

\noalign{\vspace{0.05cm}}
\hline
\noalign{\vspace{0.1cm}}
$^{\rm c}\textrm{V}_{\textrm{p,}\,{}\textrm{BH}}$ ($\textrm{km}\,{}\textrm{s}^{-1}$) & 
$0.2^{+0.3}_{-0.1}$ & $0.5^{+1.2}_{-0.3}$ & $0.8^{+2.5}_{-0.5}$ & 
$2.8^{+2.0}_{-1.5}$  &  
$0.8^{+3.0}_{-0.5}$ & $0.7^{+0.7}_{-0.5}$  \\

$^{\rm d}\langle{}\textrm{V}_{\textrm{BH}}\rangle{}$ ($\textrm{km}\,{}\textrm{s}^{-1}$) & 
 $ 0.3^{+0.8}_{-0.2}$ & $ 1.0^{+3.8}_{-0.6}$ &
$1.9^{+3.5}_{-1.1}$ & 
$4.9^{+10.9}_{-2.6}$ & 
$2.1^{+2.8}_{-1.1}$ & $0.8^{+1.4}_{-0.3}$  \\

$^{\rm e}\textrm{V}_{\textrm{rms,}\,{}\textrm{BH}}$ ($\textrm{km}\,{}\textrm{s}^{-1}$) & 
 $0.465$ & $1.51$ & $2.45$ & $6.83$ &  $2.62$ & $1.01$  \\

\noalign{\vspace{0.05cm}}
\hline
\noalign{\vspace{0.1cm}}
$^{\rm f}\textrm{e}_{\textrm{p,}\,{}\textrm{PSR-A}}$ $(\times{}10^{-5})$ & 5$^{+3}_{-5}\cdot{}10^{4}$ & 2$^{+77}_{-2}$ & 1$^{+70}_{-1}$ & 
1$^{+7}_{-1}$ & 
3$^{+76}_{-2}$ & 5$^{+744}_{-45}$ \\
\noalign{\vspace{0.05cm}}
\hline
\noalign{\vspace{0.1cm}}

$^{\rm g}\langle{}\frac{\Delta{}\textrm{E}}{\textrm{E}_\textrm{0}}\rangle{}_{\textrm{PSR-A}}$ $(\times{}10^{-3})$ & $40^{+858}_{-40}$ &  $4^{+712}_{-4}$ & $9^{+175}_{-9}$ & 
$5^{+4082}_{-5}$ & 
$2^{+577}_{-2}$ & $-1^{+6}_{-716}$ \\
\noalign{\vspace{0.05cm}}
\hline
\noalign{\vspace{0.1cm}}
$^{\rm h}\langle{}\frac{\Delta{}\textrm{E}}{\textrm{E}_\textrm{0}}\rangle{}_{\textrm{BH}}$ $(\times{}10^{-2})$ & $0.1^{+1.0}_{-0.9}$ & $0.4^{+8.9}_{-0.8}$ & $1.0^{+18.4}_{-1.1}$ & 
$3.3^{+119.0}_{-3.3}$ & 
$1.8^{+47.8}_{-1.8}$ & $0.7^{+12.5}_{-0.7}$ \\

\noalign{\vspace{0.05cm}}
\hline
\noalign{\vspace{0.1cm}}
$^{\rm i}\langle{}\frac{\Delta{}\textrm{J}}{\textrm{J}_\textrm{0}}\rangle{}_{\textrm{BH}}$ $(\times{}10^{-2})$ & $-0.01^{+1.43}_{-1.53}$ & $6^{+63}_{-27}$ & $-2^{+25}_{-15}$ & $-2^{+16}_{-12}$ & 
$-2^{+9}_{-8}$ & 
$-0.3^{+2.2}_{-2.2}$ \\

\noalign{\vspace{0.05cm}}
\hline
\noalign{\vspace{0.1cm}}
$^{\rm l}\xi_{\textrm{BH}}\,{}(\xi_{\textrm{BH, FBs}})$ & 0.3 (2.4) & 0.1 (0.9) & 0.2 (0.8) & 
0.4 (0.5) & 
0.7 (1.1) & 0.9 (1.6)\\
\noalign{\vspace{0.05cm}}
\hline
\enddata
\tablenotetext{a}{\footnotesize{Peak value of the end-state 
recoil velocity distribution of [PSR-A,CO] considering only FBs. 
$^{\rm b}$Average recoil velocity of [PSR-A,CO] considering only FBs. 
$^{\rm c}$Peak value of the end-state recoil 
velocity distribution of \bhbh binary considering FBs and IONs. 
$^{\rm d}$Average recoil velocity of \bhbh binary considering FBs and
IONs.  
$^{\rm e}$Root mean square recoil velocity of \bhbh binary considering FBs and
IONs.
$^{\rm f}$Peak value 
of the final eccentricity distribution of [PSR-A,CO], considering only FBs. 
$^{\rm g}$Relative average change of the [PSR-A,CO] binding energy
normalized 
to its initial value (E$_0$), considering only the
FBs. 
$^{\rm h}$Relative average change of the \bhbh binary binding energy 
normalized to its  initial value (E$_0$) considering FBs, IONs and Qs. 
$^{\rm i}$Relative average change of the modulus of the orbital
angular momentum of \bhbh
normalized to its  initial value ($J_0$) considering FBs, IONs and Qs. 
$^{\rm l}$$\xi{}_{\textrm{BH}}$ as defined in eq. [2] 
obtained averaging over FBs, IONs and Qs. 
$\xi{}_{\textrm{BH,FBs}}$  obtained considering only the FBs.\break
The end-state distributions of the physical quantities are
highly asymmetric. To quantify the skewness we have introduced
asymmetric values for their dispersions.
Dispersions around the mean values  are 
calculated considering those values which contain $34\%{}$ of the total area 
in the left and  right wings, respectively. Dispersion around the  
peak values are calculated considering those values 
which contain $50\%{}$ of the 
total area  descending from the peak.\\
The 12,000 binary-binary encounters simulated have been subdivided as follows: 
1000 (for the
$500\msun ,1 \msun$ binary); 1000 (50,1.4); 1000 (30,3); 1000 (10,10); 
1000 (50,10);7000 (200,10).}}
\end{deluxetable}



\begin{deluxetable}{lllllll}
\footnotesize
\tabcolsep 0.12truecm
\tablecaption{Statistics of the outgoing states in the post-recycling
scenario. 
\label{tbl-2}}
\tablewidth{0 pt}
\tablehead{
& \colhead{($500\,{}\textrm{M}_{\odot{}},\,{}1.0\,{}\textrm{M}_{\odot{}}$)}&
\colhead{($50\,{}\textrm{M}_{\odot{}},\,{}1.4\,{}\textrm{M}_{\odot{}}$)} & 
\colhead{($30\,{}\textrm{M}_{\odot{}},\,{}3\,{}\textrm{M}_{\odot{}}$)}   & 
\colhead{($10\,{}\textrm{M}_{\odot{}},\,{}10\,{}\textrm{M}_{\odot{}}$)}   &
\colhead{($50\,{}\textrm{M}_{\odot{}},\,{}10\,{}\textrm{M}_{\odot{}}$)} &
\colhead{($200\,{}\textrm{M}_{\odot{}},\,{}10\,{}\textrm{M}_{\odot{}}$)} 
}
\startdata

FBs     & 0.452 & 0.623 & 0.645 & 0.867
 &  0.657 &  0.594 \\
EXs        & 0.0 & 0.001 & 0.0 & 0.0 & 0.0 & 0.0\\
IONs        (Triplets) & 0.273 (0.267) & 0.052 (0.043) & 0.083 (0.072) & 
0.059 (0.048) &  0.107 (0.092) & 0.121 (0.103)\\
Qs              s & 0.272 & 0.313 & 0.247  & 
0.072  &  0.233 & 0.277 \\
UEs                   & 0.003  & 0.011 & 0.025  &  0.002 & 0.003 &
0.008 \\
\noalign{\vspace{0.05cm}}
\hline

\enddata

\end{deluxetable}

\newpage
\begin{deluxetable}{lll}
\footnotesize

\tablecaption{End-states in the pre-recycling scenario$^*$. \label{tbl-3}}
\tablewidth{0pt}
\tablehead{
  & \colhead{($500\,{}\textrm{M}_{\odot{}},\,{}1.0\,{}\textrm{M}_{\odot{}}$)}
  & \colhead{($50\,{}\textrm{M}_{\odot{}},\,{}10\,{}\textrm{M}_{\odot{}}$)}
 }
\startdata

$\textrm{V}_{\textrm{p,}\,{}\textrm{PSR-A}}$ ($\textrm{km}\,{}\textrm{s}^{-1}$)& $45^{+40}_{-10}$ & $25^{+90}_{-20}$\\

$\langle{}\textrm{V}_{\textrm{PSR-A}}\rangle{}$ ($\textrm{km}\,{}\textrm{s}^{-1}$)  & $62^{+43}_{-18}$ & $67^{+104}_{-34}$\\

\noalign{\vspace{0.05cm}}
\hline
\noalign{\vspace{0.1cm}}
$\textrm{V}_{\textrm{p,}\,{}\textrm{BH}}$ ($\textrm{km}\,{}\textrm{s}^{-1}$) & $0.4^{+0.3}_{-0.4}$ & $3.8^{+1.0}_{-1.5}$ \\

$\langle{}\textrm{V}_{\textrm{BH}}\rangle{}$ ($\textrm{km}\,{}\textrm{s}^{-1}$) & $0.5^{+1.3}_{-0.2}$ & 
$3.2^{+0.9}_{-2.2}$  \\

$\textrm{V}_{\textrm{rms,}\,{}\textrm{BH}}$ ($\textrm{km}\,{}\textrm{s}^{-1}$) &  $0.602$ &  $3.51$ \\
\noalign{\vspace{0.05cm}}
\hline
\noalign{\vspace{0.1cm}}

$\textrm{e}_{\textrm{p,}\,{}\textrm{PSR-A}}$ & 0.9$^{+0.1}_{-0.6}$ &  0.4$^{+0.5}_{-0.1}$\\
\noalign{\vspace{0.05cm}}
\hline
\noalign{\vspace{0.1cm}}

$\langle{}\frac{\Delta{}\textrm{E}}{\textrm{E}_\textrm{0}}\rangle{}_{\textrm{PSR-A}}$ $(\times{}10^{-1})$ & $-1.6^{+3.4}_{-1.3}$ &
$-3.5^{+1.2}_{-4.7}$ \\
\noalign{\vspace{0.05cm}}
\hline
\noalign{\vspace{0.1cm}}
$\langle{}\frac{\Delta{}\textrm{E}}{\textrm{E}_\textrm{0}}\rangle{}_{\textrm{BH}}$ $(\times{}10^{-2})$ & $0.1^{+26}_{-0.1}$ & $1^{+53}_{-1}$ \\
\noalign{\vspace{0.05cm}}
\hline
\noalign{\vspace{0.1cm}}
$\langle{}\frac{\Delta{}\textrm{J}}{\textrm{J}_\textrm{0}}\rangle{}_{\textrm{BH}}$ $(\times{}10^{-2})$ & $-0.01^{+2.37}_{-2.33}$ & $-1^{+13}_{-9}$ \\
\noalign{\vspace{0.05cm}}
\hline
\noalign{\vspace{0.1cm}}
$\xi_{\textrm{BH}}\,{}(\xi_{\textrm{BH, FBs}})$ & 0.2 (1.7) & 0.3 (1.0)\\
\noalign{\vspace{0.05cm}}
\hline

\enddata
\tablenotetext{*}{\footnotesize{Tabulated values are as in Table 1.\\
We have simulated   on a number 4000 of binary-binary encounters  for every
type of binary black holes ($\Mbh,\mbh$ in solar masses): 3000 (for the
500,1 binary); 1000 (50,10).}}
\end{deluxetable}

\newpage
\begin{deluxetable}{lll}
\footnotesize
\tablecaption{Statistics of the outgoing states in the pre-recycling scenario. \label{tbl-4}}
\tablewidth{0pt}
\tablehead{  & 
\colhead{($500\,{}\textrm{M}_{\odot{}},\,{}1.0\,{}\textrm{M}_{\odot{}}$)}
& \colhead{($50\,{}\textrm{M}_{\odot{}},\,{}10\,{}\textrm{M}_{\odot{}}$)}
}

\startdata

FBs      & 0.068 & 0.232
\\
EXs       & 0.0 & 0.0 \\
IONs        (Triplets) & 0.844 (0.669) & 0.686 (0.341)\\
Qs                & 0.087 & 0.079  \\
UEs                    & 0.0003  & 0.003\\
\noalign{\vspace{0.05cm}}
\hline

\enddata
\end{deluxetable}


\newpage
\begin{deluxetable}{lll}
\footnotesize

\tablecaption{Minimum neutron star density$^*$  \label{tbl-5}}
\tablewidth{0pt}
\tablehead{
\colhead{$\textrm{M},\,{}\textrm{m}$ ($\textrm{M}_{\odot{}}$)} & 
\colhead{$\textrm{a}_{\textrm{max,BH}}$ (A.U.)} &
\colhead{$\textrm{f}\,{}\textrm{n}_{\textrm{NS}}\,{}(\#{}/\textrm{pc}^{3})$}
}

\startdata
$500,\,{}1.0$ & 1   & 2.9$\cdot{}10^{2}$\\
$50,\,{}1.4$  & 1   & 1.5$\cdot{}10^{3}$\\
$30,\,{}3$    & 2   & 1.1$\cdot{}10^{3}$\\  
$10,\,{}10$   & 3.6 & 7.6$\cdot{}10^{2}$\\  
$50,\,{}10$   & 6   & 2.0$\cdot{}10^{2}$\\ 
$200,\,{}10$  & 7   & 5.5$\cdot{}10^{1}$\\
\enddata
\tablenotetext{*}{\footnotesize{Minimum neutron star density $fn_{\rm NS}$
in the cluster core to ensure an
interaction every billion years, i.e., over a time
comparable to the lifetime of PSR-A in the halo. 
The values of $fn_{\rm NS}$ refer to the  pre-recycling scenario  for a ($500
\msun,1\msun$) binary and to the post-recycling scenario for all remaining  binaries.
 }}
\end{deluxetable}

\end{document}